\documentclass[pre,superscriptaddress,twocolumn,showpacs]{revtex4}
\usepackage{epsfig,amsmath,amssymb,graphics,color,calc}

\newcommand{\be}{\begin{equation}}
\newcommand{\ee}{\end{equation}}
\newcommand{\ba}{\begin{eqnarray}}
\newcommand{\ea}{\end{eqnarray}}

\newcommand{\pdif}[2]{\frac{\partial #1}{\partial #2 } }

\begin{document}

\title{Spontaneous and induced dynamic correlations in glass-formers II: \\
Model calculations and comparison to numerical simulations}

\author{L.~Berthier} 
\affiliation{Laboratoire des Collo{\"\i}des, Verres
et Nanomat{\'e}riaux, UMR 5587, Universit{\'e} Montpellier II and CNRS,
34095 Montpellier, France}

\author{G.~Biroli} 
\affiliation{Service de Physique Th{{\'e}o}rique 
Orme des Merisiers -- CEA Saclay, 91191 Gif sur Yvette Cedex, France} 
 
\author{J.-P.~Bouchaud} 
\affiliation{Service de Physique de l'{\'E}tat Condens{\'e} 
Orme des Merisiers -- CEA Saclay, 91191 Gif sur Yvette Cedex, France} 
\affiliation{Science \& Finance, Capital Fund Management 
6-8 Bd Haussmann, 75009 Paris, France} 
 
\author{W. Kob} 
\affiliation{Laboratoire des Collo{\"\i}des, Verres
et Nanomat{\'e}riaux, UMR 5587, Universit{\'e} Montpellier II and CNRS,
34095 Montpellier, France}
 
\author{K. Miyazaki} 
\affiliation{Department of Chemistry, Columbia University, 3000 Broadway, 
New York, NY 10027, USA} 
 
\author{D. R. Reichman} 
\affiliation{Department of Chemistry, Columbia University, 3000 Broadway, 
New York, NY 10027, USA} 

\date{\today}

\begin{abstract}
We study in detail the predictions of various theoretical approaches,
in particular mode-coupling theory (MCT) and kinetically constrained
models (KCMs), concerning the time, temperature, and wavevector
dependence of multi-point correlation functions that quantify the
strength of both induced and spontaneous dynamical fluctuations.  We
also discuss the precise predictions of MCT concerning the statistical
ensemble and microscopic dynamics dependence of these multi-point
correlation functions.  These predictions are compared to simulations
of model fragile and strong glass-forming liquids.  Overall, MCT fares
quite well in the fragile case, in particular explaining the observed
crucial role of the statistical ensemble and microscopic dynamics,
while MCT predictions do not seem to hold 
in the strong case. KCMs provide a simplified
framework for understanding how these multi-point correlation
functions may encode dynamic correlations in glassy
materials. However, our analysis highlights important unresolved
questions concerning the application of KCMs to supercooled liquids.
\end{abstract}


% 64.70.Pf Glass transitions
% 05.20.Jj  Statistical mechanics of classical fluids 

\maketitle

\section{Introduction}
\label{introduction}
\setcounter{equation}{0}

Dynamic heterogeneity is  a well established feature of the 
behavior of a diverse class of systems close to their glass transition 
temperatures~\cite{Donth,walter,ediger,sillescu,richert,hans}.  
Given the relatively recent realization of the importance 
of dynamic heterogeneity, it is not surprising that the systematic 
characterization of such spatio-temporal behavior, and 
the lengthscales associated with it, is far from complete.  
Much recent effort has been expended to devise correlation functions that 
effectively and quantitatively probe dynamic 
heterogeneity~\cite{sharon,doliwa,yamamoto,japonais,FP,silvio2,gc,TWBBB}.  
The 
theoretical understanding of the behavior of such correlation functions is 
still in its infancy.  This is to be contrasted with our relatively mature 
understanding of bulk structure and dynamics in supercooled liquids as 
measured by simple, low order correlation functions (such as 
intermediate scattering functions) 
that can only indirectly hint at dynamic 
heterogeneity~\cite{Donth,berthier}.  

In the first paper of this series, denoted in the following as 
I~\cite{BBBKMRI},
we set out to provide a general understanding of the behavior of a 
particular class of multi-point correlation functions that encode 
information concerning the growing dynamical lengthscale in supercooled 
liquids. To set the stage for the present work we briefly recall 
some definitions and results obtained in I.
Let $f({\bf r},t) = o({\bf r},t) o({\bf r},0)$ 
be the instantaneous value of a local two-time 
correlator at position ${\bf r}$ and time $t$, and 
$[f(t)]_{\bf r} = V^{-1} \int d^d {\bf r} f({\bf r},t)$ 
its spatial average over a 
large but finite volume $V$. The thermal average $\langle [f(t)]_{\bf r} 
\rangle$ is a standard two-time correlator, such as the 
intermediate scattering 
function when the observable $o({\bf r},t)$ is the excess 
density $\rho({\bf r},t)-\rho_0$.
A previously defined multi-point susceptibility is the following 
four-point dynamic susceptibility,
\be
\chi_4(t) = N \left\langle [\delta f(t)]_{\bf r} ^2 \right\rangle 
=  \rho
\int d^d {\bf r} \left\langle 
\delta f({\bf r},t) \delta f({\bf 0},t)\right\rangle,
\label{chi4def}
\ee
where we introduced the notation $\delta X \equiv X - \langle X \rangle$
for the fluctuations of the observable $X$.
From Eq.~(\ref{chi4def}), we see that $\chi_4(t)$ 
quantifies the strength of spontaneous 
fluctuations of the dynamical behavior in supercooled liquids
by their variance.
As shown by the last term in Eq.~(\ref{chi4def}) fluctuations 
become larger if this dynamic heterogeneity becomes 
increasingly spatially correlated. Since $\chi_4(t)$ is the volume 
integral of the four-point correlator 
$S_4({\bf r},t) = \left\langle 
\delta f({\bf r},t) \delta f({\bf 0},t)\right\rangle$
(or, alternatively, in Fourier space, 
$\chi_4(t) = \lim_{{\bf q}\to 0} S_4({\bf q},t)$), 
it is directly related to the number of correlated particles, 
$\chi_4(t) \sim (\xi/a)^{d_f}$,
where $\xi$ is the dynamic correlation length, $a$ a molecular 
lengthscale, and 
$d_f$ is related to the possibly fractal 
geometry of the dynamic heterogeneity. The direct link between
$\chi_4(t)$ and the lengthscale of dynamic heterogeneity $\xi$
explains the intensity of the present experimental effort 
dedicated to its measurement~\cite{mayer,dauchot,science,agnes}.

Recognizing that spontaneous fluctuations are in general hard to 
access experimentally, we have suggested~\cite{science,BBBKMRI} to measure 
instead the response of the averaged two-time dynamical correlators 
to an infinitesimal perturbing field, 
\be
\chi_x(t) = \frac{\partial \langle [f(t)]_{\bf r}  \rangle}{\partial x}.
\label{chixdef}
\ee
In particular we have dedicated much effort to the cases 
where $x$ is either the temperature, $x=T$, or the density, $x=\rho$, 
focusing therefore on $\chi_T(t)$ and $\chi_\rho(t)$. 
In Ref.~\cite{science} it was argued that Eq.~(\ref{chixdef}) 
defines an experimentally accessible multi-point dynamic
susceptibility which is a relevant alternative to $\chi_4(t)$. 
There are two important arguments to support this claim, which we only 
summarize for $\chi_T(t)$, but they directly carry out 
to $\chi_\rho(t)$. 
The first one is that, for a classical fluid evolving
via Newton's equations at constant number of particles, $N$, volume $V$, 
and energy, $E$, the following
fluctuation-dissipation theorem holds:
\be
k_B T^2 \chi_T(t) = V \langle [\delta f(t)]_{\bf r} [\delta h(t)]_{\bf r} 
\rangle = \int d^d {\bf r} \langle 
\delta f({\bf r},t) \delta h({\bf 0},0) \rangle,
\label{fdt}
\ee
where $[e(t)]_{\bf r} = V^{-1} \int d^d {\bf r} \, e({\bf r},t)$
is the instantaneous value of the energy density, 
and $k_B$ the Boltzmann constant.
The similarity between Eqs.~(\ref{chi4def}) and (\ref{fdt}) is striking. 
The new 
susceptibility $\chi_T(t)$ quantifies the strength of correlations
between dynamic fluctuations and energy fluctuations. As
shown by the last term in Eq.~(\ref{fdt}) $\chi_T(t)$ 
becomes larger if dynamical and energy fluctuations 
become increasingly spatially correlated.
Since $\chi_T(t)$ is proportional to the volume 
integral of the three-point correlator 
$S_T({\bf r},t) = \left\langle 
\delta f({\bf r},t) \delta h({\bf 0},0)\right\rangle$
(or, alternatively, $\chi_T(t) = \lim_{{\bf q}\to0} S_T({\bf q},t)$), 
it
is also directly related to a 
correlation volume, which makes it an equally appealing quantity.
A second argument establishing the relevance of $\chi_T(t)$ is the
fact that $\chi_T(t)$ and $\chi_4(t)$ can be related by the following 
inequality:
\be 
\chi_4(t) \geq \frac{1}{c_P} T^2 \chi_T^2(t),
\label{chi4bound}
\ee 
where $c_P = V \langle [\delta h(t)]_{\bf r}^2 \rangle / (\rho T^2)$
is the constant volume specific heat 
expressed in units of $k_B$.
The result (\ref{chi4bound}) can be understood 
by formal consideration about statistical ensembles (see I),  
or more simply by noting that 
the relation (\ref{chi4bound})
stems from the fact that the (squared) cross-correlation  
between two observables (encoded in $\chi_T(t)$) cannot be larger than 
the product of their variances (encoded in $\chi_4(t)$ and $c_P$).

In I, we focused on the thermodynamic ensemble 
dependence and the dependence on the microscopic dynamics.  Using general 
theoretical arguments, we gave qualitative and quantitative 
guidelines for these dependences.  The ensemble variability of global 
multi-point indicators of dynamical heterogeneity (corresponding 
to fluctuations of intensive dynamical correlators) is not surprising, 
given what is already understood about the ensemble dependence of simpler 
susceptibilities near standard critical points~\cite{hansen,lebo}. 
Importantly, this 
ensemble dependence allows for the derivation 
of the rigorous bound (\ref{chi4bound}) 
on $\chi_4(t)$ that is potentially useful for providing a simple 
experimental estimate of the lengthscale 
associated with dynamical heterogeneity near $T_{g}$.  
That this bound becomes a good approximation for $\chi_4(t)$
above $T_{g}$ was 
checked in simulations of both 
strong and fragile glass forming liquids in I. The predicted dependence on 
the underlying nature of the dynamics is perhaps more surprising, 
especially in light of the fact that simulations of simple dynamical 
correlation functions show no such non-trivial 
dependence~\cite{gleim,szamel2,tobepisa}. Again, in I we have confirmed this 
striking prediction by atomistic simulations.

Having outlined some generic properties of a 
class of multi-point indicators of dynamical heterogeneity, 
and confirmed these basic predictions in I, we now turn to the 
information contained in specific theories of glassy dynamics.  In 
particular, we address in this paper various properties of these 
susceptibilities from the standpoint of simple mean-field spin-glass 
models~\cite{leticia}, 
the mode-coupling theory (MCT) of supercooled liquids~\cite{Gotze}, 
and kinetically constrained
models (KCMs)~\cite{reviewkcm}. 
Our choice of theoretical models is natural: to our knowledge, only 
MCT and KCMs offer a detailed theoretical description
of dynamic heterogeneity in supercooled liquids.
We aim to confront these theories with the 
general theoretical properties outlined in I, as well as with 
simulations of atomistic glass-forming systems. The outcome of this 
exercise will be a greater understanding of the successes and failures of 
these theories, which will lead us to formulate 
a number of questions related  to the comparison of 
these models with the expectations outlined in~I.

This paper is organized as follows.
In Sec.~\ref{sectionMCT}, we present the results predicted by MCT. 
This includes general scaling behavior, as well as the dynamics and
ensemble dependence as derived via a field-theoretical approach to MCT. 
Within this approach 
we can show in particular that strong ensemble and dynamics dependence 
of dynamic fluctuations arises, while no such dependence is expected
for averaged quantities~\cite{Szamel1991}. 
This section discusses the wavelength dependence of $\chi_4(t)$ 
for liquid state MCT, while 
the relationship between $\chi_4(t)$ and 
$\chi_T(t)$ within $p$-spin models for which MCT is exact
is performed in Appendix A.
In Sec.~\ref{sectionkcm} we turn to the  ensemble and dynamics dependence of
$\chi_4(t)$ and $\chi_T(t)$ in KCMs.
Here, we discuss different models with varying degrees of
cooperativity. Interesting unresolved questions, concerning the relevance of
KCMs to model molecular glasses, are outlined in this section. 
In Sec.~\ref{MD}, the prediction of these various models are compared to
atomistic simulations.
In Sec.~\ref{conclusion}, we conclude
and we detail the successes and failures of the theoretical models in
light of the comparison with simulations, and give a summary
of these comparisons in Table I. 

\section{Mode-Coupling Theory of dynamical fluctuations}
\label{sectionMCT}

\subsection{MCT and dynamic fluctuations}

Because it starts from a microscopic description of supercooled liquids
and ends up with a complete description of its dynamics, 
MCT is a powerful tool for the interpretation and prediction 
of the qualitative and  
quantitative behavior of slow dynamics in glass-forming liquids and 
colloids, at least not too close to the glass transition~\cite{Gotze}. 
The MCT transition is usually described as a small scale phenomenon,  
the self-consistent blocking of the particles in their 
local cages~\cite{Gotze}. This is surprising since on general grounds 
a diverging relaxation time is expected to arise from processes
involving an infinite number of particles (leaving aside the case of
quenched obstacles)~\cite{MS}.
Actually, the cage mechanism requires some kind of correlation in space:
in order to be blocked by one's neighbors, the neighbors themselves
must be blocked by their neighbors and so on 
until a certain scale that, intuitively, sets the relaxation timescale of the
system. The fact that within MCT ``cages'' are correlated objects \cite{BBMR} 
will in fact become
clear below.  

This ``local-cage'' point of view was challenged in the context of 
mean-field disordered 
systems by Franz and Parisi \cite{FP}; see also \cite{KT,Wolynes} for
early results.  
For these models the dynamical equations for correlators are 
formally equivalent to 
the schematic version of the MCT equations.
Franz and Parisi \cite{FP} argued that a dynamical susceptibility
similar to $\chi_4(t)$ in these 
models has a diverging peak at the dynamical mode-coupling
transition. The Franz-Parisi susceptibility 
is further discussed in Appendix A.
Although a lengthscale cannot be defined 
in mean-field models, a diverging susceptibility is the usual 
mean-field symptom for a diverging 
lengthscale  
in finite dimensions. More recently, two of the authors (BB)~\cite{BB},  
using a field-theoretical approach to MCT, clearly showed the 
existence of a diverging length within MCT and analyzed the critical 
properties of dynamical fluctuations. In that work the role of 
conserved quantities, 
emphasized in I, was overlooked. As we show 
in the following, BB's results for 
$\chi_4(t)$ are correct either for dynamics without any conserved variables
(as is the case for disordered $p$-spin systems with Langevin 
dynamics), or in ensembles where all conserved variables are fixed, 
i.e. $NVE$ for Newtonian dynamics and 
$NVT$ for Brownian or Monte-Carlo dynamics. 

When there are conserved variables the four-point correlation function 
$S_4({\bf q},t)$ can
be decomposed in two terms, in agreement with the general considerations
of I.  
These two terms reflect different physical contributions for ${\bf q}=0$:
one is the contribution in the ensemble where all conserved variables
are strictly fixed, and the 
second arises from the fluctuations of dynamically conserved variables that
feed back into the dynamical correlations.
The second term (for $q=0$)
is therefore absent in an ensemble where these variables are fixed.  This
latter term is the one that yields a lower bound for 
$\lim_{{\bf q}\rightarrow 0}S_4({\bf q},t)$, as 
expressed in Eq.~(\ref{chi4bound}).
The bound involves   
the derivative $\chi_{x}(t)$ defined in Eq.~(\ref{chixdef}), where  
$x$ is a conserved variable. For example 
$x=\rho$ for hard-spheres, where the density 
is a conserved quantity both for Brownian and Newtonian dynamics, 
or $x=H$ (or $E$), the enthalpy (or the 
energy), in cases where temperature is the relevant control parameter. 
One can of course also focus on dynamical responses with
respect to thermodynamic control parameters such as the pressure or the
temperature. One formulation is related to  
the other via a trivial thermodynamic change of variables and the chain
rule.  In the following, for simplicity,
we will always focus on the derivative with respect to conserved degrees
of freedom.

In the next subsections we shall uncover the critical properties of the
dynamical fluctuations and dynamical responses 
discussed above, and obtain and analyze 
quantitative predictions for dynamical responses within MCT. 
We numerically confirm  these results within the $p$-spin
model in \ref{appendixkuni}.

\subsection{Dynamic scaling and critical behavior}

In the following, using the field-theoretical framework developed 
in I, we obtain the 
critical behavior of dynamical fluctuations close to the MCT transition.
We focus in particular on $\chi_4(t)$, $S_4({\bf q},t)$, and $\chi_{x}(t)$.

\subsubsection{Ladder diagrams within MCT}

Different derivations of MCT follow 
a common strategy: write down exact or
phenomenological stochastic equations for the evolution of the slow conserved
degrees of freedom and then use a self-consistent one-loop approximation
to close the equations. 
For instance, in the case of Brownian dynamics the only conserved
quantity is the density, and the so-called Dean-Kawasaki 
equation~\cite{Kawasaki1994,Dean1996}
has been analyzed (see Refs.~\cite{ABL,das,MR} for a 
discussion of the different field-theories).
Field-theories are obtained through the
Martin-Siggia-Rose-deDominicis-Janssen method, where
one first introduces response
fields enforcing the
correct time evolution and then averages over the 
thermal noise~\cite{Zinn-Justin}. 

The direct derivation of MCT equations starting 
from field-theory is difficult
and different approaches have been pursued~\cite{das}. 
It is still unclear how to obtain in a consistent way 
the standard MCT equations 
derived by the Mori-Zwanzig formalism~\cite{Gotze,MR,CR}.
Indeed, if time-reversal symmetry is preserved, 
one-loop self-consistent equations 
are {\it not} the standard MCT equations,
but have similar qualitative properties~\cite{ABL}. 
They lead in particular to the same critical behavior of the correlators. 
This issue is not relevant here because 
we focus on qualitative properties of dynamic fluctuations
which depend only on the critical properties of the MCT transition. 

The starting point for describing dynamic fluctuations 
within field theory is the Legendre 
functional~\cite{Zinn-Justin,DeDominicisMartin,BlaizotRipka}
$\Gamma(\Psi_a,G_{a,b})$ (here and in the following we use 
the notations introduced in I):
\begin{equation}\label{2PI}
\Gamma(\Psi_a,G_{a,b})=-\frac{1}{2}\mbox{Tr} \log G 
+\frac{1}{2}\mbox{Tr}\, G_{0}^{-1} [G+\Psi\Psi]-
\Phi_{2PI}(\Psi,G), 
\end{equation}
where $\Phi_{2PI} (\Psi_a,G_{a,b})$ is the sum of all two-particle 
irreducible Feynman diagrams (namely those that cannot be decomposed in two
disjoint pieces by cutting two lines) constructed with the vertices of
the theory, using the full propagator $G$ as the lines and $\Psi$ as
the sources ($G_{0}$ is the bare
propagator)~\cite{Zinn-Justin,DeDominicisMartin}. The first  
derivative  of
$\Gamma(\Psi_a,G_{a,b})$ leads to the self-consistent equations for the order
parameters themselves (including $G$'s), whereas the second derivatives 
lead to the equation for the fluctuations
of the order parameters.

\begin{figure}
\psfig{file=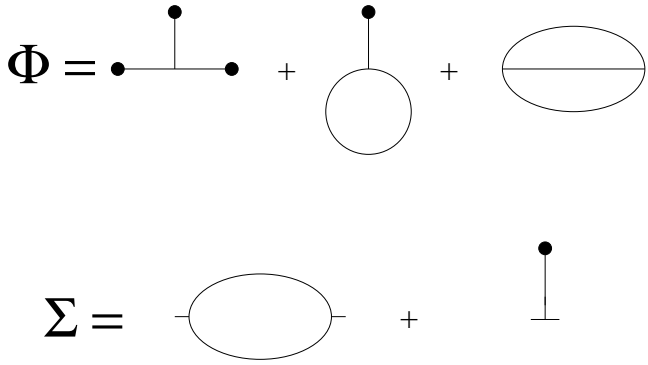,width=8.5cm}
\caption{\label{Fig1} Three diagrams approximating 
$\Phi_{2PI} (\Psi_a,G_{a,b})$ within the MCT approximation.
The resulting expression of the self-energy is also shown. 
Lines are full propagators, and
dots are conserved variables.}
\end{figure}

All field-theoretic derivations of MCT consist of one-loop self-consistent 
equations for the dynamical structure factor. At the level of the 
functional this corresponds to an approximation of 
$\Phi_{2PI} (\Psi_a,G_{a,b})$ in which only 
the first three diagrams of Fig.~\ref{Fig1} are considered.
They are constructed from a
three-leg vertex that is present in all field-theories of dense
liquids. The black dots  
represent the $\delta \Psi$ attached as 
sources and the lines the full propagators
of the theory. The corresponding expression of the self-energy 
$\Sigma=\delta \Phi/\delta G$
is also shown. Note that the second diagram is not present in the usual
expression for the self-energy
used in the MCT equations because the solution of the self-consistent
equation for $\Psi$ leads 
to $\delta \Psi=0$. 
As discussed in I, the reason
is that the average value of the response fields is zero and 
the bare values of the physical slow fields
are not corrected at any order of the self-consistent 
expansion because they correspond to conserved
variables, whose average value is not fixed by the dynamics 
but through the initial conditions.

\begin{figure}
\begin{center}
\psfig{file=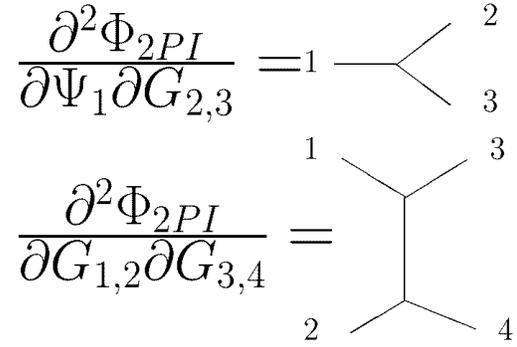,width=6.5cm}
\caption{Diagrammatic expression of 
${\delta^2 \Phi_{2PI}}/{\delta \Psi_1\delta G_{2,3}}$
and ${\delta^2 \Phi_{2PI}}/{\delta G_{1,2}\delta G_{3,4}}$ within MCT.}
\label{Fig2}
\end{center}
\end{figure}

However, when the matrix of second derivatives of $\Gamma$ is
considered, it is important to keep the second self-energy diagram
because it gives a contribution that cannot be neglected.  
The matrix of second derivatives reads:
\begin{equation}
\begin{aligned}
&
\frac{\delta^2 \Gamma}{\delta G_{1,2}\delta G_{3,4}}=
\left[G^{-1}_{1,3}G^{-1}_{2,4}-
\frac{\delta^2 \Phi_{2PI}}{\delta G_{1,2}\delta G_{3,4}}\right],\\
&
\frac{\delta^2 \Gamma}{\delta \Psi_1\delta G_{2,3}}= 
-\frac{\delta^2 \Phi_{2PI}}{\delta \Psi_1\delta G_{2,3}},
\\
&
\frac{\delta^2 \Gamma}{\delta \Psi_1\delta \Psi_2}=(G_{0}^{-1})_{1,2}.  
\end{aligned}
\end{equation}
As in I, we denote these operators respectively as $A$, $B$, and $C$. 
The diagrammatic expressions for the second derivatives of $\Phi_{2PI}$
are shown in Fig. \ref{Fig2}.    
Note that we show only the contributions that are non-zero when 
evaluated for the average quantities (in particular for $\delta \Psi=0$). 

\begin{figure}
\psfig{file=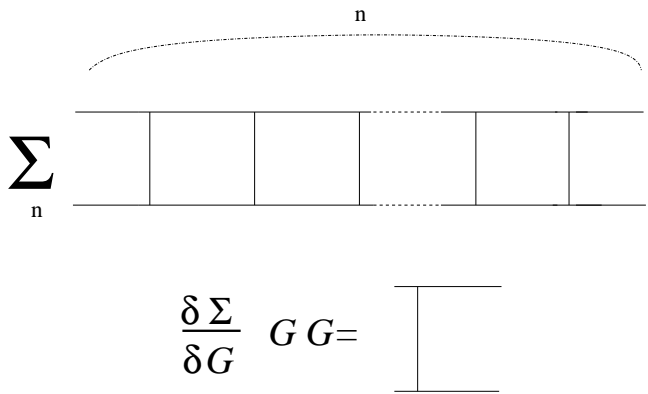,width=8.5cm}
\caption{\label{Fig3} Expression for 
$A^{-1}$ 
within MCT. It consists in a sum of $n$-ladders constructed from 
the elementary block 
$({\delta \Sigma}/{\delta G})GG$ shown in the second line, see Eq. (6-a).}
\end{figure}

All dynamic fluctuations can then be
expressed in terms of $A^{-1},B,C$ (see I). 
In particular, the four-point fluctuations 
\begin{align}
& 
\left\langle \left\{
\tilde \psi_a({\bf x},t)\tilde \psi_b({\bf x}',t')
-\Psi_a({\bf x},t)\Psi_b({\bf x}',t')
\right\} \right.
\nonumber \\
&  \,\,\, \times 
\left.\left\{
\tilde \psi_c({\bf y},s)
\tilde \psi_d({\bf y}',s')
-\Psi_c({\bf y},s)\Psi_d({\bf y}',s')
\right\}\right\rangle_c,
\end{align}
are given by
\begin{equation}\label{ABC}
A^{-1}+(A^{-1}B)\{C-B^{\dagger}A^{-1}B\}^{-1}(A^{-1}B)^{\dagger},
\end{equation}
evaluated at the matrix element 
$[a, b,$ ${\bf x},$ ${\bf x'}, t, t';~$ $c, d,{\bf y},{\bf y'}, s, s']$.
The explicit expression for $A^{-1},B,C$ makes it clear that within MCT
the critical properties of dynamical fluctuations come only from
$A^{-1}$. Indeed, $C$ is just the inverse of the bare propagator,
whereas $B$ is the bare vertex. These quantities have no critical
behavior at the MCT transition. 
Instead, using the general results of I and the MCT expression of 
${\delta^2 \Phi_{2PI}}/{\delta G\delta G}$, one finds
that $A^{-1}$ corresponds to the sum of $n$-ladder diagrams shown 
in Fig.~\ref{Fig3}.
As shown in Ref.~\cite{BB}, the resummation of these diagrams 
indeed leads to a critical contribution 
at the MCT transition. 

In particular, they give a contribution to the four point function 
$\langle \delta \rho_{-k_3}(t) \delta \rho_{k_3+q}(0)
 \delta \rho_{-k_4}(t) \delta \rho_{k_4-q}(0)\rangle$ that scales 
as~\cite{BB,BBMR}:
\begin{equation}\label{ld1}
\begin{aligned}
&
\frac{1}{\sqrt{\epsilon}+q^2}g_{\beta}\left(\frac{q^2}{\sqrt{\epsilon}},
\frac{t}{\tau_{\beta}}\right), \qquad t\propto \tau_{\beta},
\\
&
\frac{1}{\sqrt{\epsilon}[\sqrt{\epsilon}+q^2]}
g_{\alpha}\left(\frac{t}{\tau_{\alpha}}\right), \qquad t\propto \tau_{\alpha}.
\end{aligned}
\end{equation}
Note that here and in the following we use the standard 
MCT notation~\cite{Gotze,CR}. 
In particular $\epsilon=|x_c-x|/x_c$ is 
the reduced distance from the critical
point at $x=x_c$.  The function
$g_{\beta}\left(q^2/\sqrt{\epsilon},t/\tau_{\beta}\right)$
behaves as $\left(t/\tau_{\beta}\right)^a$ and
$\left(t/\tau_{\beta}\right)^b$ for small and large value of  
$\left(t/\tau_{\beta}\right)$, respectively. 
Furthermore the function 
$g_{\alpha}\left(t/\tau_{\alpha}\right)$ behaves
as $(t/\tau_{\alpha})^b$ for small values of 
$(t/\tau_{\alpha})$, see also Eqs.~(\ref{earlybeta},
\ref{earlyalpha}) below. 

Below we show that the critical behavior that emerges from ladder 
diagrams underlies all of the critical properties encoded in the
dynamical fluctuations and responses, as discussed in general terms in I.

\subsubsection{Dynamical responses}

Dynamical response functions, defined in Eq.~(\ref{chixdef}), are
particularly interesting because they provide, 
through inequalities like Eq.~(\ref{chi4bound}), an estimate of  the
relevant dynamical fluctuations, and because they are related
to three-point dynamical correlations, Eq.~(\ref{fdt}).

\begin{figure}
\psfig{file=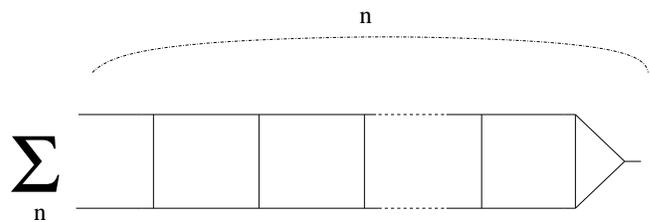,width=8.5cm}
\caption{\label{Fig4} The dynamical response, obtained from 
Eq. (\ref{responseMCT}) by noting that the inversion 
of $A=\partial_{GG}^2 \Gamma$ involves resumming ladders that 
close on $\partial_{G\Psi}^2 \Gamma$.}
\end{figure}

An exact expression for dynamical response functions can be derived
noting that $G$ is obtained by setting $\partial \Gamma/\partial G \equiv 0$. 
Differentiating this relation with respect to a conserved variable $\Psi$ 
one can easily derive the relation (see Eq.~(60) of I): 
\begin{equation}\label{responseMCT}
\chi_{\Psi}=\frac{\partial G}{\partial \Psi}
=-\left[\frac{\partial^2\Gamma}{\partial G \partial G} 
\right]^{-1}\frac{\partial^2 \Gamma}{\partial G \partial \Psi}=A^{-1}B, 
\end{equation}
which is represented in Fig.~\ref{Fig4} using MCT diagrams.

Since the derivative is taken with respect
to the average value of one of the conserved degrees of
freedom $\Psi$, the wavevector 
entering into the ladder diagrams is zero.
As a consequence, the scaling of dynamical response functions is 
given by Eq.~(\ref{ld1}), setting $q=0$:
\begin{equation}\label{dr1}
\begin{aligned}
&
\sim \frac{B(q=0)}{\sqrt{\epsilon}}\, g_{\beta}
\left(\frac{t}{\tau_{\beta}}\right), 
\qquad t\propto \tau_{\beta}, 
\\
&
\sim \frac{B(q=0)}{\epsilon}\, g_{\alpha}\left(\frac{t}{\tau_{\alpha}}\right),
\qquad t\propto \tau_{\alpha}, 
\end{aligned}
\end{equation}
where we have dropped the first argument of $g_\beta$, equal to zero here, and 
$B(q=0)$ reminds us that there is an additional contribution from 
the vertex $B$.
We show below, see Eqs.~(\ref{alpharegime}, \ref{betaregime}, \ref{earlybeta}, 
\ref{earlyalpha}), that
these results can alternatively be obtained analytically 
using standard MCT results.
However, the field-theoretical derivation shows
more clearly the role of the ladder diagrams, and is 
crucial to understand the relationship between dynamical response and 
dynamic fluctuations. 
We note that from the diagrammatic expression for $\chi_{\Psi}$ 
a clear relationship with three-point dynamical correlators appears. 
The diagrammatic expression of the correlation 
between the fluctuation of the 
dynamical structure factor and the fluctuation of a conserved 
variables $\Psi$ reads (see I):
$-A^{-1}B \{C-B^{\dagger}A^{-1}B\}^{-1}$,
which contains the same diagrams as 
$\chi_{\Psi}$, with a propagator attached 
at the end~\footnote{Note that $\{C-B^{\dagger}A^{-1}B\}$ 
is the propagator of the theory but
it coincides with $G$, the solution of the self-consistent equation, only 
if all diagrams are retained. Within the MCT approximation they are
 quantitatively different but qualitatively similar. The self-energy
 $B^{\dagger}A^{-1}B$ contains ladder diagrams but they are harmless
 because the times on the same side of the ladders coincide since they
 are attached to the same vertex. Therefore, they are related to the
 dynamical  fluctuations at equal time which are not critical.}.

In order to probe the spatial dependence of dynamical
fluctuations related to ladder diagrams, zero-wavector response
functions such as $\chi_\Psi(t)$ are not sufficient, and one should consider
instead the response to a spatially modulated external field~\cite{BBMR}.  
It was recently proved in Ref.~\cite{BBMR}
that such a $q$-dependent dynamical response function has the same scaling 
as the one anticipated from ladder diagrams in Eq.~(\ref{ld1}).   

\subsubsection{Ensemble and microscopic dynamics dependence of fluctuations}

In the following we illustrate, within MCT, the dependence on statistical
ensembles and on microscopic dynamics of dynamical fluctuations which we have
discussed in full generality in I. In particular, one finds that 
although $S_4({\bf q},t)$ and its $q \rightarrow 0$ 
limit are ensemble independent quantities, 
$\chi_4(t)=S_4({\bf q}=0, t)$ 
does depend on the ensemble and on microscopic dynamics. 
This fact reflects 
the subtle nature of global fluctuations when the thermodynamic limit
is taken~\cite{lebo}.

\begin{figure}
\psfig{file=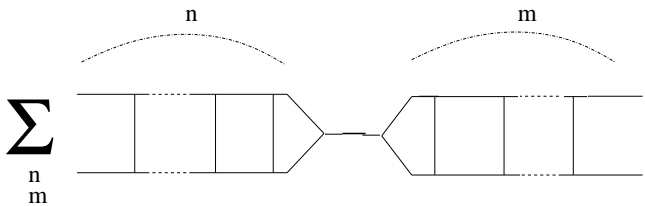,width=8.5cm}
\caption{\label{laddersquare} Representation in term of diagrams of 
$(A^{-1}B)\{C-B^{\dagger}A^{-1}B\}^{-1}(A^{-1}B)^{\dagger}$, see 
Eq.~(\ref{ABC}), within
MCT. It corresponds, roughly speaking, to 'squaring' the 
ladders of Fig. (\ref{Fig3}).}
\end{figure}

Applying the general theory developed in I one finds that
$S_4({\bf q},t)$ is given by the ladder diagrams in Fig.~\ref{Fig3} plus
``squared ladders'' as shown in Fig.~\ref{laddersquare}. 
The ladder diagrams 
shown in  Fig.~\ref{laddersquare} are joined by a propagator at 
wavector ${\bf q}$. 
In ensembles where all conserved degrees of freedom are fixed,
e.g. Newtonian dynamics in the $NVE$ ensemble or Brownian dynamics in the
$NVT$ ensemble, the propagator evaluated at $q=0$ 
vanishes, because conserved quantities do not
fluctuate on the scale of the system size (and all propagators 
related to response fields are zero because
they are proportional to $q$ at small $q$). Therefore, in 
these ensembles, simple ladder diagrams provide the sole
contribution to $\chi_4(t)$ within MCT, causing $\chi_4(t)$ to scale as
in Eq.~(\ref{ld1}) evaluated at $q=0$. This is also
true for $p$-spin models, see Appendix A.  

Instead, in ensembles where at least one conserved degree
of freedom is allowed to fluctuate, e.g. the $NVT$ or $NPT$ 
ensembles for Newtonian
dynamics, or the $NPT$ ensemble for Brownian dynamics, the 
propagator joining the
ladders in  Fig.~\ref{laddersquare} does not vanish, 
and contributes only to 
non-critical prefactors. In this case the diagrams 
corresponding to ``squared
ladder'' diagrams dominate, at least close enough to the 
transition. Note that their overall scale 
might however be small, for example if the compressibility or 
the specific heat are large or if the distance to the critical 
point becomes large. 
They lead to a modified critical behavior for $\chi_4(t)$ 
within MCT, reading: 
\begin{equation}\label{ls1}
\begin{aligned}
&
\frac{1}{\epsilon}\, \tilde{g}_{\beta}\left(\frac{t}{\tau_{\beta}}\right), 
\qquad t\propto \tau_{\beta}, 
\\
&
\frac{1}{\epsilon^2}\, \tilde{g}_{\alpha}\left(\frac{t}{\tau_{\alpha}}\right), 
\qquad t\propto \tau_{\alpha},
\end{aligned}
\end{equation}
where $\tilde{g}_{\alpha} \propto {g}_{\alpha}^2$ and 
$\tilde{g}_{\beta} \propto {g}_{\beta}^2$.
 
These results provide non-trivial relationship between dynamical 
fluctuations
in different ensembles and different microscopic dynamics. 
Although already suggested by the general theory developed in I 
they become sharp
statements within MCT. For instance, the predicted dynamic scaling 
of $\chi_4(t)$ in the $NVT$ ensemble for Newtonian dynamics is that of the 
``squared ladder'' diagrams of Eq.~(\ref{ls1}), whereas the predicted
scaling of $\chi_4(t)$ in the $NVT$ ensemble for
Brownian dynamics is that of Eq.~(\ref{dr1}), which coincides 
with the one expected for Newtonian dynamics in the $NVE$ ensemble. 
Note that the critical mechanism  underlying the dynamical fluctuations 
is the same for all 
ensembles and dynamics 
and is uniquely encoded in ladder 
diagrams. Indeed there is a unique lengthscale, diverging in the
same way for all microscopic dynamics, which underlies the 
critical behavior. However, the 
coupling to conserved degrees of freedom may produce a large 
amplification of global fluctuations. 

\subsubsection{Behavior of $S_4({\bf q},t)$ and upper critical dimension}

The behavior of $S_4({\bf q},t)$ for $q \neq 0$ is apparently simpler 
because it does not depend on the statistical ensemble.
Furthermore, all types of diagrams are present 
in its expression so that its qualitative critical behavior is
independent of the microscopic dynamics, provided that  
at least density is locally conserved. However, since 
$S_4({\bf q},t)$ contains
the two terms discussed previously (ladders and squared ladders) 
a crossover behavior might be expected.
Although the squared ladders should dominate very close to $T_c$ 
they might become sub-dominant far from the critical point.
Therefore one should be very cautious when comparing
the present MCT predictions to the behavior of real liquids where 
the mode-coupling singularity $T_c$ is
replaced by a smooth crossover towards 
an activated regime. 
In order to judge the relative importance of the two terms (ladder and
squared ladder),
one may focus on their $q=0$ value, which corresponds to $\chi_4^{NVE}$
for ladders and to $k_B T^2 \chi_T^2/c_V$ for squared ladders. 
For example, in the case of the
LJ mixture studied in I, the latter term 
becomes dominant only close to the transition $T\simeq 0.47$. As a
consequence, for higher temperatures, the contribution
of the squared ladders can
be neglected and $S_4(q,t)$
will have the behavior presented in Eq.~(\ref{ld1}).
A similar crossover is expected for $\chi_4(t)$, 
as confirmed numerically in I. The important difference with 
$S_4({\bf q},t)$ is that it is possible, at least in numerical simulations,
to disentangle the different contributions to $\chi_4(t)$ 
by working in different ensembles.

Although four-point correlators were originally hoped to be suited to 
quantify precisely dynamical heterogeneities in glass-formers, 
our results show that, although containing 
useful information on dynamical heterogeneities, they 
mix it with other, less interesting physical effects. 
 This is a further motivation to study 
dynamical response to spatially modulated fields 
introduced in~\cite{BBMR}, for which
only the simple ladder diagrams contribute and which, therefore,
allows one to obtain clearer and more direct information on 
dynamic correlations.
In future work, it would be extremely interesting to compare 
this response function computed within MCT~\cite{BBMR} to its 
direct numerical evaluation in a simulated liquid. 

Finally we note that, within MCT, the scaling of dynamic 
fluctuations in the ensemble where all conserved
variables fluctuate is different from the one
predicted by BB~\cite{BB}. This implies that the upper critical 
dimension of the theory, found
to be $d_c=6$ in~\cite{BB}, has to be revised accordingly.
Focusing on the $\beta$-regime, the fluctuations of the 
non-ergodicity parameter in a region of size $\xi\sim \epsilon^{-1/4}$ grow as 
$\delta q \propto \xi^{4-d/2}$, where $d$ is the space dimension. 
Imposing, in the spirit of a Ginzburg criterion, that $\delta q$
must be much smaller than the critical behavior of 
the order parameter,
i.e. $q_c - q \sim \sqrt{\epsilon}$, one finds 
that fluctuations become dominant below the upper critical dimension
$d_c=8$. In \cite{decoupling} it is shown   that this result can be 
obtained from diagrammatic considerations:
below $d_c=8$, corrections to MCT are found to diverge in 
the infrared regime. 

\subsection{The $k$-dependence of dynamical fluctuations within MCT}

Several different definitions of $\chi_{4}(t)$ have been employed in the 
literature~\cite{silvio2,TWBBB,dauchot,BB,glotzer,lacevic,steve,berthier,Ck}.  
Regardless of definition, since two-point density 
fluctuations must depend on the dynamically probed lengthscale 
($\sim k^{-1}$), the detailed behavior of $\chi_{4}(t)$ will 
also depend on this  lengthscale, see e.g.~\cite{Ck}. 
Physically, the dependence of the fluctuations on 
lengthscale reflects the coupling or sensitivity of cooperative motion to 
behavior on the scale of the measured two-point fluctuations. For example, the 
expectation that high-frequency phonons do not couple strongly to the 
large lengthscale dynamic heterogeneity
is reflected by the fact that a $\chi_{4}(t)$ that focuses on 
short lengthscales associated with vibrations cannot exhibit 
the sizable normalized peak values that are connected to large cooperative 
lengths (but see the discussion in \cite{TWBBB}).  
Only at a critical point would one expect all modes 
to couple in such a way that the behavior of $\chi_{4}(t)$ would 
exhibit truly universal  properties. Since this issue cannot be discussed
within $p$-spin models which contain no lengthscale
(see \ref{appendixkuni}), we turn instead in this 
section to liquid
state MCT which contains the complete wavevector dependences of dynamic
functions.

The dependence of $\chi_{4}(t)$ as a function of
lengthscale was first discussed by Glotzer and coworkers~\cite{lacevic},
who  used the definition 
\begin{equation}
\chi_{4}(t) 
=\frac{\beta V}{N^{2}}\left[ 
\langle Q^{2}(t)\rangle -\langle Q(t)\rangle^{2}
\right],
\end{equation}
where $N$, $V$ and $\beta$ are the number of particles, the volume and the 
inverse temperature, respectively, and 
$Q(t)=\sum_{ij}w(|{\bf r}_{i}(0)-{\bf r}_{j}(t)|)$ 
and 
$w(|{\bf r}_{1}-{\bf r}_{2}|) =1$ 
for 
$|{\bf r}_{1}-{\bf r}_{2}| \leq a$ 
and is zero otherwise. 
Here $a$ is a cutoff parameter. 
La\v{c}evi\'{c} {\em et al.} 
explicitly showed that within this definition of $\chi_{4}(t)$, the value 
of $a$ that maximizes the peak height is close to the global Debye-Waller 
amplitude of the mean-square displacement~\cite{lacevic}.  
For values of the cutoff that 
are larger or smaller, the absolute amplitude of $\chi_{4}(t)$ decreases.  
It can be noted in this work that the shape of $\chi_{4}(t)$ is sensitive 
to the value of $a$ as well, although no systematic study of this 
dependence was investigated. 

Using a different definition of $\chi_{4}(t)$, namely that
defined by Eq.~(\ref{chi4lj}) below, 
Dauchot {\em et al.} noted that for a weakly sheared 
granular system, the slope of $\chi_{4}(t)$ increases as the wavevector 
decreases~\cite{dauchot}.  
This particular dependence has been studied in 
more detail in a recent work~\cite{Ck}.  
From both molecular dynamics simulation and the direct analysis of a class 
of kinetic facilitated models, Chandler {\em et al.} have detailed the 
lengthscale dependence of a variety of definitions of $\chi_{4}(t)$, 
and have argued that a generic feature of this dependence is that the 
growth of $\chi_{4}(t)$ to its peak becomes significantly more rapid as 
the intrinsic lengthscale increases. It was also 
argued~\cite{Ck} that this result is inconsistent with the predictions of 
mode-coupling theory outlined in Ref.~\cite{TWBBB}.  
Since the $k$-dependent $\chi_{4}(t)$ in facilitated models has been 
discussed in detail in Ref.~\cite{Ck}, we focus below on the predictions 
of mode-coupling theory. In particular, we show that, despite statements 
to the contrary, mode-coupling theory is at least in qualitative accord 
with the behavior found from computer simulation, as detailed in
Ref.~\cite{Ck}. 

An important aspect of the physical content of the $k$-dependence of 
$\chi_{4}(t)$ is embodied in the consideration of the distinction between 
$\beta$- and $\alpha$-relaxation, and the implication that 
this distinction holds  
for the lengthscales of dynamic heterogeneity.  At a given density and 
temperature, a well-defined plateau in the two-point density correlator
for wavevectors near the first diffraction peak in 
the static structure factor $S(k)$ will 
saturate as $k$ is decreased. Eventually, as $k$ is decreased 
further the hydrodynamic regime is reached, where the local cooperative 
processes associated with dynamic heterogeneity are averaged out.
At a fixed distance from 
the dynamical transition temperature $T_{c}$, as $k$ is decreased first 
the $\beta$-relaxation window decreases in duration~\cite{fuchs1991}.  
Concomitantly, the stretching exponent of the $\alpha$-relaxation
increases continuously.  Eventually, the crossover is complete when the 
$\beta$-window is no longer observable, and the stretching exponent saturates 
at unity, signifying long-time hydrodynamic behavior.

The theoretical considerations made in Ref.~\cite{TWBBB} are based on 
the asymptotic predictions of mode-coupling theory.  Arbitrarily close 
to $T_{c}$ and in systems for which $T_{c}$ is not avoided, the 
predictions of Ref.~\cite{TWBBB} are nearly universal in 
the sense that the effects mentioned above are only seen in the strict 
$k \rightarrow 0$ limit.  By considering the $k$-dependence of induced 
susceptibility $\chi_{x}(t)$ for a fixed, finite distance from 
$T_{c}$ one gains a qualitative understanding of how the universal features 
expected for $k$ near the first diffraction peak of $S(k)$ are changed as 
$k$ decreases.

As discussed in the previous sections dynamical fluctuations encoded  
in ladder diagrams are visible either in $\chi_4(t)$ or in the dynamical 
response. 
However, dynamical responses are accessible to direct quantitative
numerical computations that are an essential tool in order to discuss
the crossover issues discussed above. 
Let us now, for completeness and clarity, 
rederive the results for dynamical responses using only standard MCT
results~\cite{Gotze} being particularly careful about the
$k$-dependence. 

In the $\alpha$-regime close to the transition, 
but still in the liquid phase,
the dynamical structure factor scales like
$F(k,t)\simeq f^{k}_{\alpha}(t/\tau_{\alpha}(\epsilon))$ where 
$\tau_{\alpha}=\epsilon^{-\gamma}$ 
and $\gamma=1/2a+1/2b$. 
Thus, we find that in the $\alpha$-regime $\chi_{x}(k, t) = \partial 
F(k,t)/ \partial x$ reads:
\begin{equation}
\label{alpharegime}
\chi_{x}(k,t)\simeq \frac{1}{\epsilon}g^{k}_{\alpha}(t/\tau_{\alpha}), 
\qquad g^{k}_{\alpha}(x)=-\frac{x}{\gamma}\frac{df^{k}_{\alpha}}{dx}.
\end{equation}
We temporarily change notation to emphasize the $k$-dependence, namely,
we promote $\chi_x(t)$ to $\chi_x(k,t)$. 
In the $\beta$-regime and close to the transition 
$F(k,t)\simeq S(k)q(k)+ S(k) h(k) \sqrt{\epsilon}f_{\beta}(t/\tau_{\beta})$ 
where $\tau_{\beta}=
\epsilon^{-1/2a}$, $q(k)$ is the non-ergodic parameter and $h(k)$ is the 
critical amplitude.
Thus, we find that in the $\beta$-regime $\chi_{x}(k,t)$ reads:
\begin{equation}\label{betaregime}
\begin{aligned}
\chi_{x}(k,t) 
& \simeq  
\frac{h(k)S(k)}{\sqrt{\epsilon}}g_{\beta}(t/\tau_{\beta}), 
\\
g_{\beta}(x) 
& 
= -f_{\beta}(x)-\frac{x}{2a} f_{\beta}'(x).
\end{aligned}
\end{equation}
Analyzing the $\beta$-regime with a large but not diverging time and
matching the $\alpha$-regime with the $\beta$-regime imposes constraints on
the large and small $x$ behavior of $g_{\beta}(x)$ and
$g^{k}_{\alpha}(x)$. Requiring that in the early $\beta$-regime the
$\epsilon$ dependence should drop out of $\chi_{x}(k,t)$, we find:  
\begin{equation}
\label{earlybeta}
\chi_{x}(k,t) \sim t^a \qquad 1 \ll t \ll \tau_{\beta}.
\end{equation}
Analogously, matching the $\alpha$- and $\beta$-regimes leads to:
\begin{equation}
\chi_{x}(k,t)\sim \epsilon^{(b-a)/2a} t^b \qquad 
\tau_{\beta}\ll t \ll \tau_{\alpha},
\label{earlyalpha}
\end{equation}
interpolating between $\chi_{x} \sim \epsilon^{-1/2}$ for $t=\tau_{\beta}$ and 
$\chi_{x} \sim \epsilon^{-1}$ for $t=\tau_{\alpha}$, 
before decaying back to zero for $t \gg \tau_{\alpha}$.

\begin{figure}
\psfig{file=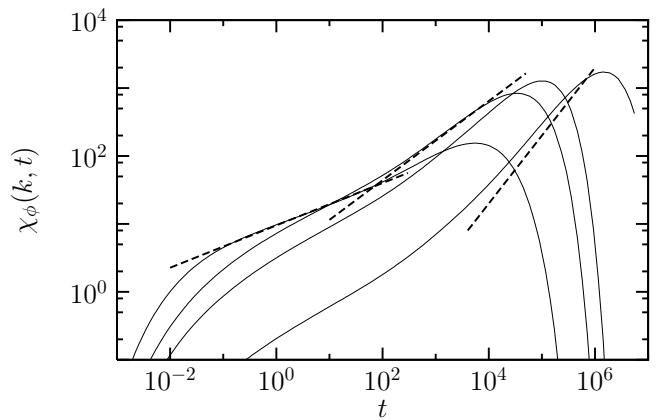,width=8.5cm}
\caption{\label{figMCTHS}
Dynamic susceptibility $\chi_\phi(k,t)$, Eq.~(\ref{chiphi}), predicted 
by MCT for hard spheres at fixed volume fraction
above the glass transition, $\phi_c - \phi = 10^{-3}$ 
for various wavevectors from
$k=19.35$ to $k=0.75$ (in particle size units)
from left to right.
Power laws for the largest $k$ are the asymptotic results 
$t^a$ and $t^b$ with $a=0.312$ and $b=0.583$, while
$\chi_\phi (t) \sim t$ describes well the data at small $k$ (rightmost curve).}
\end{figure}

All these results are valid close enough to the transition but, 
as discussed previously, 
we expect crossovers as a functions of $k$ and time. 
In order to study this issue numerically,
we solve the full, wavevector-dependent mode-coupling equations for 
the self-intermediate function $F_{s}(k,t)$ for a dense colloidal
suspension, which are directly coupled to the collective density  
fluctuations $F(k,t)$ as 
\begin{equation}
\begin{aligned}
\frac{\partial F_{s}(k,t)}{\partial t}
&
+D_{0}k^{2}F_{s}(k,t)
\\
&
+\int_{0}^{t}\!\! dt'~ 
M_{s}(k,t-t')\frac{\partial F_{s}(k,t')}{\partial t'}=0,
\end{aligned}
\end{equation}
where $D_{0}$ is the bare diffusion constant, and $M_{s}(k,t)$ is the 
self-memory function that can be expressed as
\begin{equation}
M_{s}(k,t)=\frac{\rho_0 D_{0}}{(2\pi)^{3}}\int d{\bf k}' 
\left\{\hat{\bf k}\cdot{\bf k}'c(k')\right\}^{2}
F_{s}(|{\bf k}-{\bf k}'|,t)F(k',t).
\end{equation}
Here,  $\rho_0=N/V$ is the number density, $\hat{\bf k}={\bf k}/|{\bf k}|$, 
$c(k)$ is the direct correlation function, 
$\rho_0 c(k)=1-{1}/{S(k)}$.
This equation must be solved simultaneously and self-consistently with
mode-coupling equations for the full density fluctuations
\begin{equation}
\frac{\partial F(k,t)}{\partial t}
+\frac{D_{0}k^{2}}{S(k)}F(k,t)
+\int_{0}^{t}\!\! dt'~M(k,t-t')\frac{\partial F(k,t')}{\partial t'}
=0,
\end{equation}
where 
\begin{equation}
M(k,t)=
\frac{\rho_0D_{0}}{(2\pi)^{3}}
\int\!\! d{\bf k}'~ |V({\bf k},{\bf k}')|^{2} 
F(|{\bf k}-{\bf k}'|,t)F(k',t),
\end{equation}
and 
$V({\bf k},{\bf k}')
= \hat{\bf k}\cdot{\bf k}'c(k')
 +\hat{\bf k}\cdot({\bf k}-{\bf k}')c(|{\bf k}-{\bf k}'|)$.
These equations are solved for a model hard-sphere suspension with input 
from $S(k)$ calculated from the Percus-Yevick closure
at various volume fraction $\phi$. 
The wavevector cutoff is 
taken to be $k_{c}=50$ in units of the particle size, and the grid number 
is taken to be 100.  The equations of motion are integrated with the 
algorithm of Fuchs {\em et al.}~\cite{fuchs1991}.
The induced susceptibility 
\be
\chi_{\phi}(k,t)=\frac{\partial F_{s}(k,t)}{\partial \phi},
\label{chiphi}
\ee
is computed via numerical 
differentiation.  In Fig.~\ref{figMCTHS} 
the induced susceptibility is shown for different  
values of wavevector from those higher than the first peak in $S(k)$ to 
those significantly below.  
The behavior of $\chi_{\phi}(k,t)$ for $k$ close to the first 
diffraction peak displays the two power law regimes 
described in Sec.~\ref{sectionMCT}. 
When $k$ is decreased the power law 
describing how $\chi_\phi(k,t)$ reaches its peak 
clearly shows an increasing value.
This behavior is qualitatively compatible with the one 
discussed theoretically and found in
numerical simulations of KCM's and atomistic liquids in~\cite{Ck} 
and experiments on granular materials~\cite{dauchot}. 
It makes clear 
that corrections to the critical behavior are different depending on $k$. 
In particular, the limit $k\rightarrow 0$ and $T\rightarrow T_c$ clearly do
not commute.

Note that, although the critical behavior is expected to be the
same for $\chi_{\phi}(k,t)$ and $\chi_{4}(t)$ in the $NVE$ ensemble, 
the corrections to the critical scaling might not be the same.
However, the same qualitative considerations 
regarding time and wavevector dependences should hold.
Furthermore, using the bounds derived in I the present results
yield direct quantitative predictions for the behavior of the dynamic 
susceptibility $\chi_4(t)$ in the hard sphere system since 
its scaling behavior is the same as to the one of 
$\chi_\phi^2(k,t)$, which obviously follows 
from Eqs.~(\ref{alpharegime}, \ref{betaregime}, \ref{earlybeta}, 
\ref{earlyalpha}).

\section{Dynamic susceptibility and diverging lengthscales in KCMs}
\label{sectionkcm}

\subsection{Models and observables}

In this section we proceed with the computation of the dynamic susceptibility
$\chi_T(t)$ and its comparison to the previously studied 
$\chi_4(t)$ in kinetically constrained
models (KCMs)~\cite{reviewkcm}.
Our motivation here stems from the fact that the study of
KCMs has greatly contributed to our present understanding
of the dynamically heterogeneous dynamics of supercooled 
liquids~\cite{gc,TWBBB,reviewkcm,sellitto,steve,Ck,jackle,FA,harrowell,douchebag2,BG,TBF}.
Moreover, the variety of available models allows one to
grasp the variety of possible behaviors that could possibly be
encountered in real materials. Finally the
relative simplicity of the models
makes them suitable to large scale
numerical simulations, which might help data analysis
in real materials, while scaling laws and exact
results can be obtained by standard theoretical
tools of statistical mechanics.

KCMs are spin models (lattice gas versions also exist~\cite{KA2})
generically described
by a simple, usually non-interacting Hamiltonian, and a set of
dynamic rules with non-trivial constraints forbidding some of the
transitions and therefore making the overall dynamics glassy.
In the following we will focus on spin models
characterized by the Fredrickson-Andersen Hamiltonian~\cite{FA},
\be
H = \sum_{i=1}^N n_i,
\label{FAham}
\ee
where $n_i = 0$, 1 is a binary variable defined on each point of
a hypercubic lattice in dimension $d$.
Physically, $n_i=0$ ($n_i=1$)
represents a site $i$ which is immobile (mobile), and has therefore
an energy which is smaller (larger) than the average energy,
given by
\be
\langle n_i \rangle = c(T) = \frac{1}{1+e^{1/T}}.
\ee
The spins evolve with a single spin flip dynamics, so that the model dynamics
is entirely defined
by the transition rates between states 1 and 0,
\be
n_i=0
\begin{array}{c}
\xrightarrow{~ ~ ~{\cal C}_i ~ c ~ ~ } \\ \xleftarrow[{\cal C}_i ~
(1-c)]{} \\
\end{array}
n_i=1 ,
\ee
where ${\cal C}_i$ is a kinetic constraint on site $i$ which
can become 0 depending on the local environment of site $i$,
therefore prohibiting some specific transitions.
We shall study in detail
two different spin facilitated models where the kinetic constraint
takes the following forms,
\be
{\cal C}_i = 1-\prod_j (1-n_j),
\label{1fa}
\ee
and
\be
{\cal C}_i = 1 - \prod_{j,k} (1-n_j n_k) ,
\label{2fa}
\ee
for 1- and 2-spin facilitated models, respectively. In the expressions for
${\cal C}_i$ the products are over nearest neighbors of site $i$. The
constraints
respectively become equal to 1 when, respectively, 
at least 1 or 2 of their
nearest neighbors is mobile, therefore capturing the idea of dynamic
facilitation: mobile regions locally favor the creation of more
mobility~\cite{gc,FA}.

Due to the presence of a heat bath, the dynamics of KCMs
do not conserve energy. Physically this means
that heat can be locally provided to a spin to allow the creation of
a mobility excitation without the need to borrow energy from
the neighboring sites.
In principle the results obtained from KCMs should then be compared
to the $NVT$ dynamics of molecular liquids.
As opposed to the MCT results described above, no prediction
can be made from KCMs concerning the role of a conservation law
for the energy. For kinetically constrained lattice gases, 
however, a quantitative 
comparison between spontaneous fluctuations, $\chi_4(t)$, 
and fluctuations induced by a change of density, $\chi_\rho(t)$,
can be performed~\cite{Ck}.

A second important consequence of the presence of a
heat bath is that neither
the fluctuation-dissipation relation in Eq.~(\ref{fdt}) nor the inequality
Eq.~(\ref{chi4bound}) apply to KCMs, and we are therefore left with 
three independent dynamic quantities, namely
\begin{equation}
\begin{aligned}
&
\chi_T (t) = \frac{\partial \langle P(t) \rangle}{\partial T},
\\
&
C_{PE} (t) = N \langle \delta P(t) \delta e(0) \rangle,
\\
&
\chi_4(t) = N \langle \delta P^2(t) \rangle,
\end{aligned}
\label{CPE}
\end{equation}
where we have defined the instantaneous value of the
energy density,
\be
e(t) = \frac{1}{N} \sum_{i=1}^N n_i(t).
\ee
Following earlier works on KCMs we choose to work with the persistence
function as the relevant two-time dynamical object,
\be
P(t) = \frac{1}{N} \sum_{i=1}^N P_i(t),
\ee
where $P_i(t)$ denotes the persistence of the spin $i$ between
times 0 and $t$. Its thermodynamic average, $\langle P(t) \rangle$,
has recently been the subject of a number of theoretical 
studies~\cite{gc,steve,BG,steve2,rob,cristina}.
Note that for KCMs, the Cauchy-Schwarz inequality 
could still be of some use. For the present variables this leads to
\be
\chi_4(t) \ge \frac{C_{PE}^2(t)}{T^2 c_V},
\label{CS}
\ee
where $c_V (T) = dc/dT = e^{1/T}/ [T^2 (1+e^{1/T})^2]$ is the specific
heat. The main difference between the inequalities (\ref{CS}) and
(\ref{chi4bound}) is that the right hand side of (\ref{CS})
is given by a correlation function which is not easily
accessible in experiments, contrary to the susceptibility
$\chi_T(t)$ appearing in (\ref{chi4bound}). Of course,
$C_{PE}(t)$ can be measured in numerical experiments, as
shown below.

At the level of the
spatial correlations, two distinct correlators also need
to be studied,
\ba
S_T(r,t) &=& \langle \delta P_i(t) \delta n_{i+r}(0) \rangle,
\label{CT} \\   
S_4(r,t) &=& \langle \delta P_i(t) \delta P_{i+r}(t) \rangle.
\label{C4}
\ea
Their Fourier transforms, $S_4(q,t)$ and $S_T(q,t)$, 
can equivalently
be studied, and it is obvious that $S_4(q=0,t) = \chi_4(t)$ and
$S_T(q=0,t) = C_{PE}(t)$, these quantities representing
the volume integrals
of the spatial correlations $S_4(r,t)$ and $S_T(r,t)$, respectively.

\subsection{Results for 1-spin facilitated FA models}
\label{1dkcm}

The one-spin facilitated FA
model has been studied numerically and analytically
in various spatial dimensions in
much detail~\cite{gc,TWBBB,reviewkcm,steve,Ck,BG,steve2,rob,cristina}. 
These studies have shown that
the model exhibits dynamic heterogeneity and large
spontaneous fluctuations of the two-time dynamics, although
relaxation timescales grow only in an Arrhenius fashion as
temperature is decreased,
\be
\tau_\alpha \sim c^{-\Delta} \sim \exp(\Delta / T),
\label{fatime}
\ee
with $\Delta = 3$ for $d=1$ and $\Delta = 2$ for $d>2$.
Interestingly, these works suggest that even strong material
should display dynamic heterogeneity. 
This was confirmed by simulations~\cite{bks_sim} and 
experiments~\cite{Geyer1996} which reported
deviations from the Stokes-Einstein
relation, although the FA model itself 
presents no such deviations for $d \geq 2$. 
 
As usual, the four-point
susceptibility $\chi_4(t)$ is found to have non-monotonic time dependence.
Therefore it shows a peak, $\chi_4^\star(T) =
\chi_4(t \sim \tau_\alpha)$,
whose position shifts to larger times
and whose height increases when temperature decreases. One finds
dynamic scaling~\cite{gc,steve,steve2},
\be
\chi_4^\star \sim c^{-\gamma} \sim \exp(\gamma /T) \sim
\tau_\alpha^{\gamma/\Delta},
\ee
with $\gamma = 1$ in all spatial dimensions.
The corresponding spatial dynamic
correlations have also been studied.
Analytically, one can compute these quantities approximately by making the
assumption that the system can be described as an assembly of
defects which diffuse
independently with diffusion constant $D=c$. This
was called ``independent defect
approximation'' in Ref.~\cite{TWBBB}.
In three dimensions, one finds
\be
S_4(q,t) \approx \chi_4(t) {\cal S}_4 [ q^2 \xi_4^2(t) ],
\label{S4}
\ee
with a diffusively growing lengthscale,
\be
\xi_4(t) = \sqrt{ct},
\label{diff4}
\ee
and the scaling function
\be
{\cal S}_4(x) = 2 \, \frac{x-1+e^{-x}}{x^2}.
\label{S4res}
\ee
Additionally the four-point dynamic susceptibility behaves as follows,
\be
\chi_4(t) \approx \frac{c_2}{2 c} \left( \frac{t}{\tau_\alpha} \right)^2
\exp \left( - \frac{2t}{\tau_\alpha} \right),
\ee
with $c_2$ a numerical factor.
These predictions are in good agreement with direct 
simulations of the FA model, the only discrepancy being that
the scaling function for $S_4(q,t)$ shows deviations
from its $1/q^2$ predicted large $q$ 
behavior when times become very large,
$t \gg \tau_\alpha$.

The computation of $\chi_T(t)$ is easy given that the average
persistence obeys time temperature superposition, $\langle P(t)
\rangle = f(t/\tau_\alpha)$, the scaling function $f(x)$ being well
described, for times which are not too long~\cite{cristina}, 
by a stretched exponential form, $f(x) = \exp(-x^\beta)$,
with $\beta=1/2$ for $d=1$ and $\beta=1$ for $d>2$. Therefore one
immediately 
gets, 
\be \chi_T (t) = - \frac{\Delta \beta}{T^2} \left(
\frac{t}{\tau_\alpha} \right)^\beta \exp \left[ - \left(
\frac{t}{\tau_\alpha} \right)^\beta \right]. 
\label{chit} 
\ee 
This
shows that $\chi_T(t)$ displays a non-monotonic time dependence with
a peak arising at time $t \sim \tau_\alpha$, diverging as
$\chi_T^\star \sim - 1/T^2$ when $T$ goes to zero. Finally, 
the behavior of $\chi_T(t)$ before the peak, $t \ll \tau_\alpha$, 
is a power law, $\chi_T(t) \sim t^\beta$, $\beta$ being the value of the 
stretching exponent characterizing also the $\alpha$-relaxation. 

If one considers the quantity $T^2 \chi_T^2 / c_V$ appearing
in the inequality (\ref{chi4bound}), one finds at the peak,
\be \frac{T^2}{c_V} (\chi_T^\star)^2  \sim c^{-1} \sim \exp(1/T)
\sim \chi_4^\star,
\label{miracle}
\ee
so that both sides of the inequality
(\ref{chi4bound}) have similar scaling
properties at low temperatures in this model. This is not an obvious result
given that these quantities are not related by the thermodynamic relations
and inequalities outlined in Sec.~\ref{introduction}.

Notice, however, that this similarity appears coincidental because
the whole divergence of the first term
in Eq.~(\ref{miracle}) is due to
the very strong temperature dependence of the specific heat
at low temperature which itself results from the
non-interacting FA Hamiltonian (\ref{FAham}). In real materials,
the specific heat is almost temperature independent when the glass
transition is approached and the growth of the term $T^2 \chi_T^2 /c_V$
is mainly due to the growing susceptibility $\chi_T(t)$ itself.

Following steps similar to those described in Ref.~\cite{TWBBB} it
is possible to compute both the correlator in 
Eq.~(\ref{CT}) and its volume integral in 
Eq.~(\ref{CPE}) within the independent defect approximation. 
In three dimensions, one finds
for $S_T(q,t)$ a scaling form very similar to Eq.~(\ref{S4}), 
\be 
S_T(q,t)
\approx C_{PE}(t) {\cal S}_T [q^2 \xi_T^2(t)] ,
\ee 
with 
\be
C_{PE}(t) \approx c_1 \left( \frac{t}{\tau_\alpha} \right) \exp
\left( - \frac{c_1 t}{\tau_\alpha} \right), \label{cpe} \ee where
$c_1$ is a numerical factor, the corresponding correlation
lengthscale 
\be 
\xi_T(t) = \sqrt{ct} ,
\label{diffT} 
\ee 
and the scaling function 
\be 
{\cal S}_T(x) = \frac{1-e^{-x^2}}{x^2}.
\label{STres} 
\ee 
These calculations show that, within 1-spin
facilitated models, the physical content of the correlators $S_4$
and $S_T$ is essentially the same. Physically, this is because two
sites are dynamically correlated, and therefore contribute to $S_4(r,t)$,
if they are visited by the same diffusing mobility defect. Similarly, two
sites contribute to $S_T(r,t)$ if one of them contains at time 0 the
first defect which will visit the second one for $t>0$. This implies
that the correlation lengthscales $\xi_4$ and $\xi_T$ both reflect
the simple activated diffusion of point defects, and therefore
contain the same physical information; Eqs.~(\ref{diff4}) and
(\ref{diffT}) show that they are indeed equal. Additionally, the
spatial correlators $S_4(q,t)$ and $S_T(q,t)$ are found to differ in their
detailed expression, Eqs.~(\ref{S4res}, \ref{STres}), but they
have the same asymptotic behaviors, $S_T(q \xi_T \ll 1) \sim const$
and $S_T(q \xi_T \gg 1) \sim 1/q^2$, reminiscent of an
Ornstein-Zernike form.

An additional piece of information derived from Eqs.~(\ref{chit}) and 
(\ref{cpe}) is the similar time dependence and scaling with
temperature found for the quantities $T^2 \chi_T(t)$ and $C_{PE}(t)$,
despite the fact a fluctuation-dissipation relation such as
Eq.~(\ref{fdt}) does not hold.
Numerically we indeed find that both terms quantitatively differ, 
although merely by a numerical factor. 

\begin{figure}
\psfig{file=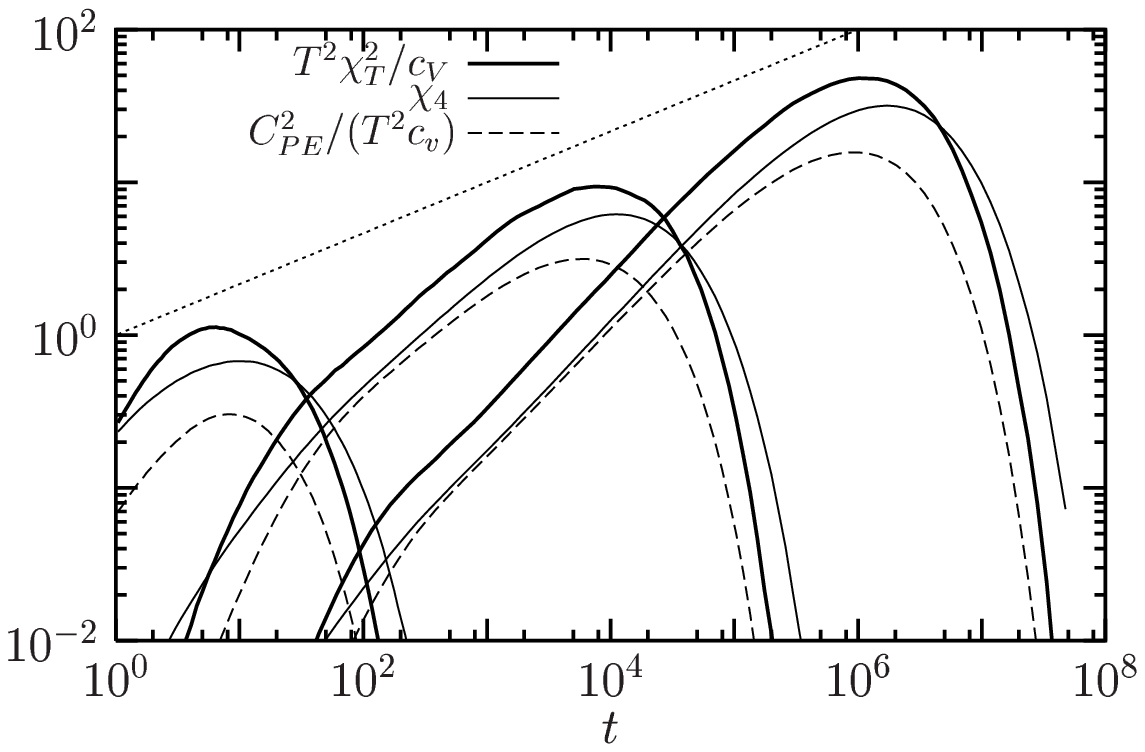,width=8.5cm}
\psfig{file=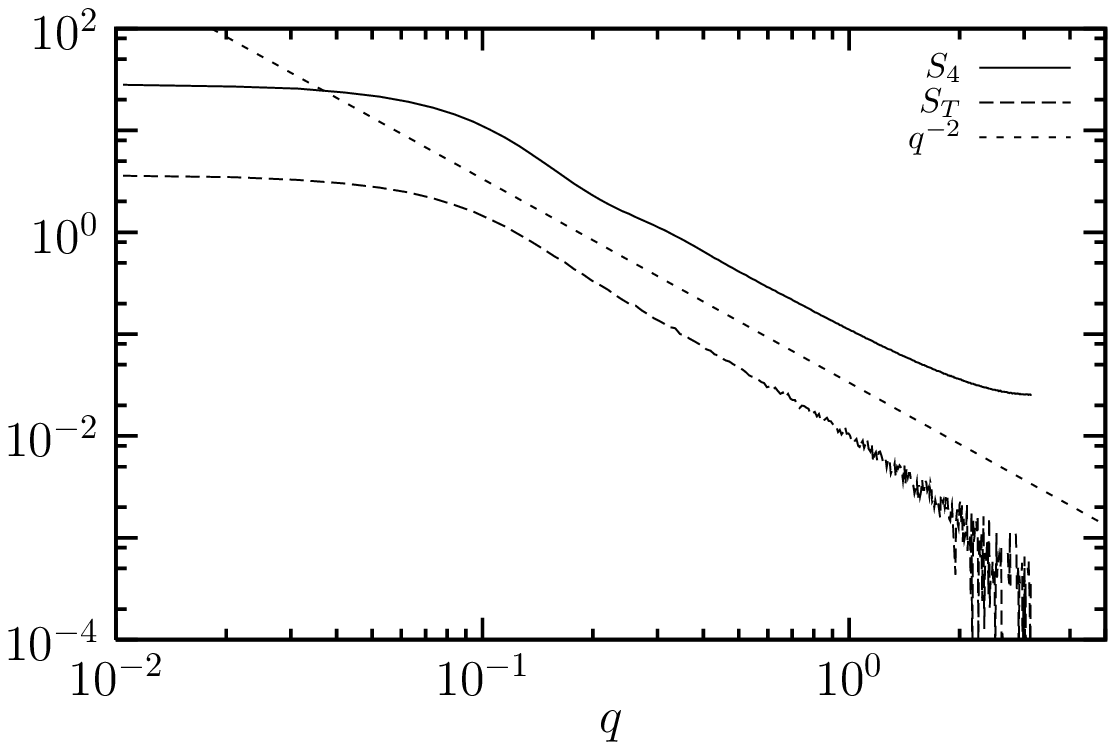,width=8.5cm}
\caption{\label{figFA} 
Dynamic susceptibilities and spatial correlations
in the one-spin facilitated FA model in one dimension.
Top: Various dynamical susceptibilities are shown
as a function of time for temperatures
$T=1.0$, 0.5 and 0.2 (from left to right). They behave similarly with time,
and their peak scale as $\tau_\alpha^{\gamma/\Delta} = \tau_\alpha^{1/3}$,
as shown with dots.
Bottom: Dynamic structure factors $S_4(q,t)$ and
$S_T(q,t)$ at time $t=\tau_\alpha$ for $T=0.3$. Both functions
behave as a constant at small $q$, and as $1/q^2$ at large wavevectors,
as shown with a dashed line.}
\end{figure}

The independent defect approximation is thought to be a good
representation of the 1-spin facilitated model above its critical
dimension, $d> 2$, as confirmed by our numerical simulations in
$d=3$. 
In Fig.~\ref{figFA},
we provide additional numerical evidence that these  findings
are also correct in $d=1$ as well. The top figure shows that
the time dependence and scaling with temperature of three different
quantities, $\chi_4(t)$, $C_{PE}^2(t)/(T^2 c_V)$ and $T^2 \chi_T^2(t)
/c_V$ are the same. Moreover, the Cauchy-Schwarz inequality
(\ref{CS}) is satisfied by our numerics, at it should be, while the
inequality (\ref{chi4bound}) derived for Newtonian dynamics is found to
be violated, by a factor which is about 2 at the peak.
The bottom panel in Fig.~\ref{figFA} shows $S_4(q,\tau_\alpha)$ and
$S_T(q,\tau_\alpha)$ for the FA model in $d=1$.
As predicted from the independent defect approximation,
both correlation functions are slightly different in their shape
but share a common behavior, a plateau at small $q$
and a $1/q^2$ decay at large $q$. We have
checked that the equivalence $\xi_4 (t) \sim \xi_T(t)$ also
holds in numerical simulations, confirming that
dynamic-dynamic and dynamic-energy spatial correlations are 
essentially  equivalent quantities in the context of non-cooperative
KCMs.
This is physically expected since
by definition of the kinetic constraints in (\ref{1fa}), it is
those regions with high potential energy which trigger the dynamics
of the nearby sites: this is the essence of the dynamic 
facilitation idea. 

We conclude this section on one-spin facilitated
models by briefly discussing the case of the
East model~\cite{jackle}, which is defined with the same FA Hamiltonian
(\ref{FAham}) and is also a 1-spin facilitated model where the
kinetic constraint is defined similarly to  Eq.~(\ref{1fa}), the 
only difference
being that the product appearing in (\ref{1fa}) is now restricted
to only one neighbor in each spatial direction. This ``hyper''-East model
was called the North-or-East-or-Front (NEF) model in $d=3$~\cite{nef}.
This directionality of the constraint
makes the dynamics of the East model slower than that of the FA model,
and relaxation timescales now grow in a super-Arrhenius fashion, so that
the exponent $\Delta$ appearing in (\ref{fatime}) becomes
temperature dependent, $\Delta(T) \sim - \ln c(T)$.
The East model is therefore a KCM  for fragile glasses.
Additionally, time temperature superposition does not hold. Relaxation is
still described by stretched exponentials but the stretching exponent
is also temperature dependent, with $\beta(T) \sim T$ at low 
temperature~\cite{juanpe,peter}.
Despite these qualitative differences
between strong and fragile models, our main conclusions
still hold. The three dynamic susceptibilities shown in Fig.~\ref{figFA}
also track each other, and this is again the result of subtle
compensations between the scaling of correlations functions and
the strong temperature dependence of the defect concentration.
Similarly, the two different lengthscales
$\xi_4$ and $\xi_T$ also bear the same physical content,
although they now grow sub-diffusively with time~\cite{TWBBB}.
This subdiffusive behavior affects the approach 
of the dynamic susceptibilities to their maximum. In the $d=1$ East model, 
one finds that before the peak $\chi_T(t) \sim t^{b(T)}$, 
where the exponent $b(T) \approx \beta(T)$ 
should decrease slowly when $T$ decreases. We find
numerical values $b \approx 0.2 - 0.4$ in the time window
of our Monte Carlo simulations, where relaxation timescales 
increase from $\tau_\alpha \sim 10^4$ to $\tau_\alpha \sim 10^8$.  

\subsection{Results for a 2-spin facilitated FA model}

\label{22kcm}

By comparison with one-spin facilitated models, much less is known
about the behavior of 2-spin facilitated models, because relaxation
does not proceed by activated diffusion (or even sub-diffusion)
of point defects~\cite{FA,harrowell}.
In some cases, asymptotic mechanisms have been described
which show that relaxation timescales grow very rapidly when
temperature is decreased, although no finite temperature
divergence is found~\cite{harrowell,TBF}.
In these mechanisms, relaxation occurs via the diffusion of
``super-defects'' whose concentration decreases when
$T$ decreases and whose size is itself an increasing
function of temperature. For this reason these models are sometimes
called ``cooperative KCMs''.
Very recently, a  KCM was specifically engineered
to yield an example of a finite temperature singularity, but we do not
discuss this example further~\cite{stupid_adhoc_model}.

An additional point of theoretical interest of cooperative KCMs is that, when
studied on Bethe lattices, they display a dynamical transition
at finite  temperature which is reminiscent of the mode-coupling
singularity described in Sec.~\ref{sectionMCT}. 
Moreover, dynamic fluctuations
can be studied in some analytic detail in the Bethe limit, while
no analytic study of dynamic fluctuations on finite dimensional
lattices is available.

We now focus on the 2-spin facilitated model in dimension $d=2$, the
``22FA model'', as a specific example of a cooperative model. Our
choice is motivated by the relatively large number of earlier
studies dedicated to this model~\cite{FA,harrowell,TBF}, 
the fact that its Bethe
limit was also considered~\cite{sellitto}, and that it is sufficiently far
from its mean-field limit that deviations from mean-field behavior
are clearly observed. It was indeed  realized early on that the
model does not display a power law divergence of its relaxation time
at finite $T$~\cite{FA,harrowell}, contrary to more constrained models
where numerics seemed to indicate the presence of a mean-field like
singularity~\cite{KA2}, now discarded~\cite{TBF,fss}.

Adapting the general results of Ref.~\cite{TBF} to the specific
example of the 22FA model, we expect the following scaling results. 
The relaxation time grows as \be \tau_\alpha  \sim \exp \left(
\frac{a}{c} \right) \sim \exp \left[ a \exp  \left( \frac{1}{T}
\right) \right], \label{texpexp} \ee where $a$ is a numerical
factor. The double exponential divergence makes the 22FA a very
fragile glass-former model.

The scaling of the four-point dynamic susceptibility is obtained as
follows. At a given temperature, relaxation occurs via the diffusion
of super-defects of size $\ell(T)$. By coarse-graining the system up
to size $\ell$, relaxation then resembles the diffusion of
independent defects, and the results of the independent defect
approximation can be carried out. Therefore, we expect $\chi_4^\star \sim
c_\ell^{-1}$, where $c_\ell$ is the concentration of super-defects.
Using the results of Ref.~\cite{TBF}, we get
\be 
\chi_4^\star
\sim  \exp \left( \frac{a}{c} \right) \sim \tau_\alpha.
\label{expexp} 
\ee
We evaluate the leading divergence of $\chi_T(t)$ by assuming time
temperature superposition, i.e. $\chi_T(t) = \partial
f(t/\tau_\alpha)/\partial T$. Using (\ref{texpexp}) we get $\chi_T^\star \sim
\exp ( a/c )/(T^2 c)$, up to an irrelevant numerical prefactor. As a
consequence, the right hand side of the inequality (\ref{chi4bound})
scales as 
\be 
\frac{T^2}{c_V} (\chi_T^\star)^2 \sim c^{-3} 
\sim (\ln\tau_\alpha )^3. 
\label{log} \ee
By comparing Eqs.~(\ref{log}) and (\ref{expexp}), we
conclude that the dynamic heterogeneity quantified through
$\chi_4(t)$ and $T^2 \chi_T^2(t)/c_V$ are very different, since
$\chi_4(t)$ is predicted to diverge as a power of $\tau_\alpha$, while
the term involving $\chi_T$ 
should diverge only logarithmically with $\tau_\alpha$. For cooperative
models, the ``coincidental'' compensation due to the specific heat
arising in non-cooperative model is not effective.

Since these results are expected to hold only very close to $T=0$,
we have performed numerical simulations of the 22FA model. In these
Monte Carlo simulations, we cover the temperature regime $T=2.6$ down
to $T=0.43$, which corresponds to about 7 decades of relaxation
timescales. In this temperature window, $\tau_\alpha$ cannot be
fitted with an inverse power law $\tau_\alpha \sim
(T-T_c)^{-\alpha}$ as in the Bethe limit, showing that
strong
non-mean-field effects are indeed present. However, the form
(\ref{texpexp}) is not completely successful either, suggesting that
the true asymptotic regime is beyond the realm of numerical
simulations (see \cite{Dawson} for a discussion of this point 
in a similar context), 
and that the numerical regime lies somewhat
in a crossover regime.

\begin{figure}
\psfig{file=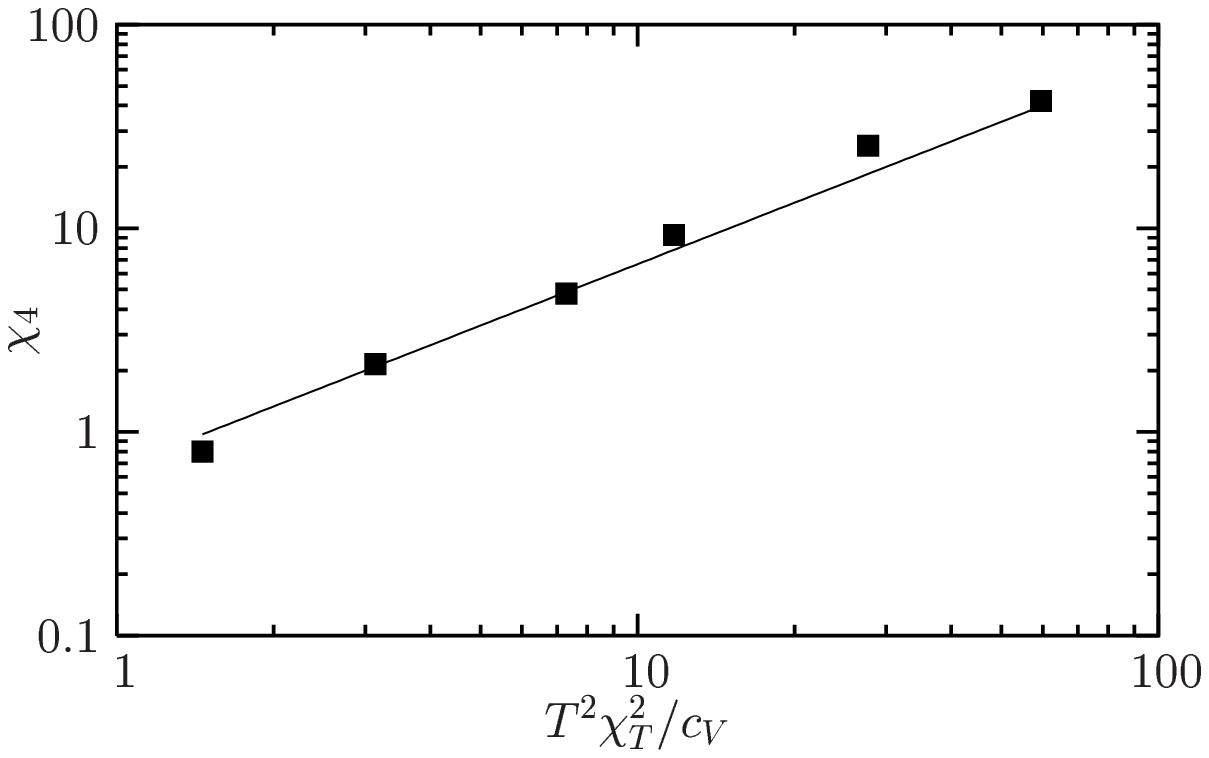,width=8.5cm}
\psfig{file=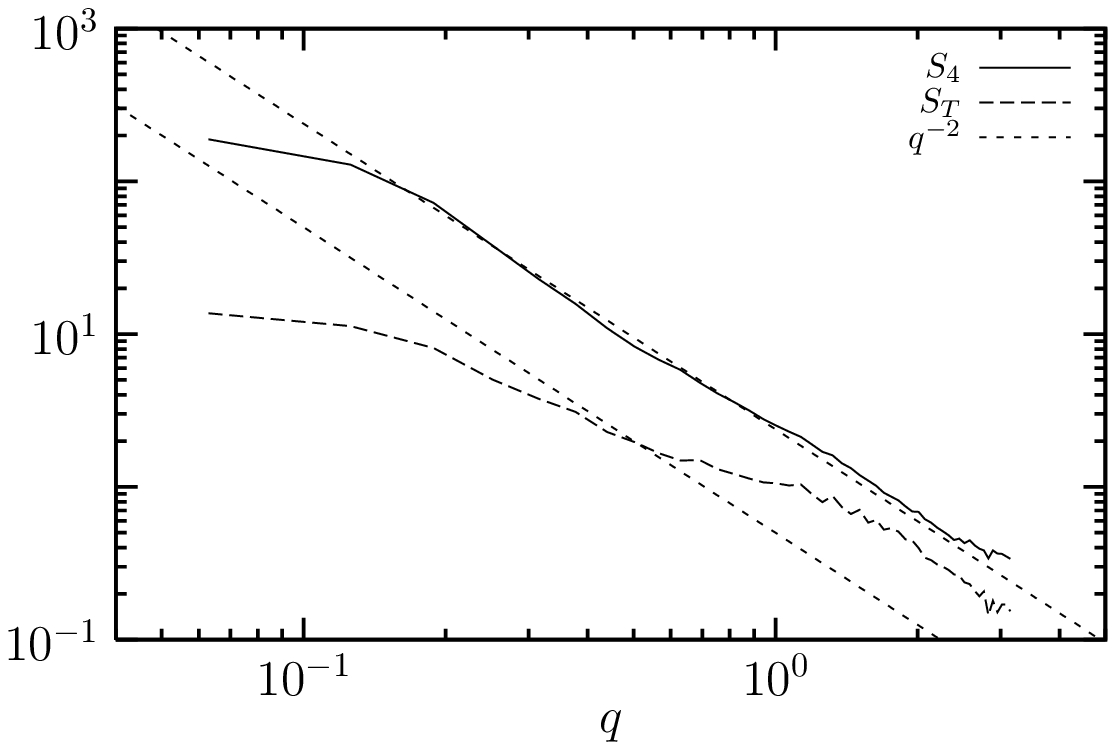,width=8.5cm} 
\caption{\label{22FA} Dynamic
susceptibilities and spatial 
correlations in the 2-spin facilitated FA model
in two dimensions. Top: Comparison between the peak values of
$\chi_4$ and $T^2 \chi_T^2/c_V$ for different temperatures covering
about 7 decades in relaxation timescales. The full line represents
the proportionality between both quantities. Bottom: Dynamic
structure factors $S_4(q,t)$ and $S_T(q,t)$ at time $t=\tau_\alpha
\sim 10^6$ for $T=0.428$. Both functions behave as a constant at small $q$, but
have different large $q$ behaviors since the $1/q^2$ dashed line is
consistent with $S_4$ only.}
\end{figure}

In Fig.~\ref{22FA} we compare the evolution of the peak of
$\chi_4(t)$ and the corresponding peak in $T^2 \chi_T^2(t) / c_V$
for the entire temperature range we have been able to 
access numerically. 
Quite strikingly we find that both functions scale very similarly on the
whole temperature range. 
A similar result was recently reported for 
a cooperative constrained lattice gas in two-dimensions~\cite{Ck}.
This similarity holds also at the level of
the whole time dependence (not shown). From numerical simulations
only, we would therefore conclude that the coincidence between the
two terms already found for non-cooperative models also applies in
cooperative models. This numerical evidence is contradicted by the
asymptotic analytic arguments given above. A possible way to
reconcile these results is to assume that the temperature regime we
have studied in the simulations 
is still too close to the mean-field Bethe lattice
limit, where the scaling  $\chi_4 \sim T^2 \chi_T^2/ c_V$ is indeed
expected to hold. This argument is however clearly weakened by the
fact that many observables (timescales, persistence functions, and
others, see Ref.~\cite{prepa}) show visible deviations from their mean-field
limit in the same temperature regime.

In the bottom panel of Fig.~\ref{22FA},
we also show the comparison of the spatial correlations
(\ref{CT}) and (\ref{C4}) measured in Fourier space.
Whereas both correlators were found to be very similar
for non-cooperative models, numerics clearly reveals that
the shapes of the dynamic structure factors
$S_4$ and $S_T$ differ. While $S_4\sim 1/q^2$ seems to hold
at large wavevectors, $q \xi_4 \gg 1$,
we find a different behavior for $S_T$, namely $S_T \sim 1/q^{1.3}$. Note that
this fit is not very satisfactory, revealing a
more complex structure of this correlator, possibly
related to the presence of two lengthscales in the model: the size
of the super-defects, and the typical distance separating them.
We conclude that dynamic-dynamic and
dynamic-energy correlations might contain 
slightly distinct physical information in cooperative KCMs.  
This is physically expected because
an isolated defect, which represents a positive
local fluctuation of the energy, cannot diffuse and relax the
neighboring sites. Therefore the correspondence
between energy fluctuations and dynamical fluctuations is not one-to-one
as in, e.g., the 1-spin FA model. By this argument one can predict that
$S_T(r,t) < S_4(r,t)$ at small $r$, and therefore
a faster initial decay of $S_T(r,t)$ with $r$. 
In Fourier space, this means
a slower large $q$ decay of $S_T(q,t)$ than that of $S_4(q,t)$, as
observed in Fig.~\ref{22FA}. 

Nevertheless, a dynamic correlation lengthscale
can be defined from both $S_4(q,t)$ and $S_T(q,t)$ as the inverse of 
the wavevector above which structure factors start to decay. The data
shown in Fig.~\ref{22FA} clearly indicate that these two lengthscales
are very close. A possible  
interpretation is that despite their complex structure,
super-defects remain  associated with some positive energy fluctuations, so
that the lengthscale  
extracted from three- and four-point functions could indeed
be equivalent, as in the case of non-cooperative models. 
A similar situation was encountered 
in our atomistic simulations in I.

\subsection{Remarks and open questions on 
ensemble and dynamics dependence and KCMs}

We have studied in the context of 
kinetically constrained
spin models (KCMs) the dynamic 
susceptibility $\chi_T(t)$ and the associated three-point 
dynamics-energy spatial correlations, and their 
link with the more standard four-point susceptibility 
$\chi_4(t)$. 
Although  the thermodynamic relations derived in I 
for supercooled liquids do not hold for kinetically constrained 
spin models  (because energy is not dynamically conserved), they
seem to be approximately valid.

The underlying reason is that in non-cooperative KCMs
the energy fluctuations that are important for the dynamics
are effectively conserved because of the kinetic constraint. 
This is clearer in the example of 
the 1-spin facilitated FA model where
a facilitating spin can 
disappear only by annihilation 
with another facilitating spin. Similarly, a facilitating spin 
can be created only by branching from another facilitating spin.
But these two processes happen very rarely (see 
Refs.~\cite{steve,steve2,rob} for a 
detailed analysis and discussion of timescales).
Therefore, the main relaxation mechanism
is diffusion of the facilitating regions (energy fluctuations) which are 
conserved in an effective way, as assumed  in the independent defect
approximation.

Comparing our results for KCMs to the general theoretical
considerations of I opens interesting issues
related to the applicability of KCMs to supercooled liquids. 
Since dynamical fluctuations strongly 
depend on statistical ensembles and microscopic 
dynamics, this immediately raises important questions: 
\begin{itemize}
\item For which ensemble are the dynamical fluctuations 
of real liquids supposed to be described by KCMs? 
\item What type of liquid dynamics should one choose to compare real
dynamical fluctuations to the prediction of KCMs?  
\end{itemize}

These questions are clearly related to the 
coarse-graining procedure
that is often invoked~\cite{FA,arrow,BG}, 
but never truly performed, to map real liquids 
to KCMs. Were this procedure known, the answer to the previous questions 
would be clear. Unfortunately, this formidable
task has not yet been accomplished.
On the other hand, our results show that
this issue is important if one wants to 
compare KCM predictions for dynamic susceptibilities $\chi_4$
and $\chi_T$ to experimental and numerical 
results on realistic models.

For kinetically constrained spin models, the
answer to the first of the above questions seems fairly easy 
even without the coarse graining procedure. 
Only in the most general ensemble where
all conserved quantities fluctuate does one have
$\lim_{q\rightarrow 0} S_4(q,t)=S_4(0,t)$. 
Since this equality holds in KCMs, we 
conclude that KCMs should apply to real liquids in the most general 
statistical ensemble, i.e. $NPT$ for most practical purposes.

The second question is instead much more subtle. 
From a general point of view since there is no conserved
quantities in spin models, KCMs could be thought as representative
of a dynamics without conserved quantities. 
Of course all physical dynamics should  at 
least conserve density. However, if one considers
Brownian dynamics for supercooled liquids for 
which temperature is the relevant control parameter, while 
density plays a minor role (see section II.E.3 in I), 
it might be reasonable to expect that density fluctuations
do not couple strongly to dynamical fluctuations. 
One is then tempted to conclude that KCMs are models of 
real liquids with Brownian 
or stochastic dynamics.

However, this tentative answer is contradicted by 
several facts.
First, real supercooled liquids obviously evolve with Newtonian dynamics. 
Second, we just discovered that 
the inequality (\ref{CS}) provides a good approximation to 
$\chi_4(t)$ for KCMs. A similar result holds 
for liquids with Newtonian dynamics in the $NPT$ ensemble (see I)
but {\it not} for liquids with stochastic dynamics~\cite{szamel}.

Taking the opposite view that KCMs represent, for some unclear
reason, liquids with Newtonian dynamics is also unsatisfactory
because the saturation of the inequality
(\ref{CS}) in KCMs is principally due to the behavior of the specific heat 
that decreases exponentially fast 
as temperature decreases. But a very small specific heat 
is incompatible with experimental measurements 
of the thermodynamics of supercooled liquids~\cite{BBT}. 
Correcting for this fact as in Ref.~\cite{CG} then leads 
to poor estimates of $\chi_4(t)$ via dynamic response functions,
in disagreement with atomistic simulations~\cite{science,BBBKMRI}.

The case of kinetically constrained lattice gases is less problematic 
if taken as models of glass/jamming transition in hard sphere systems, rather
than molecular liquids. In this case, the only conserved quantity that 
matters is the 
density and therefore 
there are no ambiguities since density is conserved both 
in kinetic lattice gases and in real systems.    

KCMs provide a natural mechanism 
explaining correlations between 
energy fluctuations and dynamic heterogeneity. However, in order to 
compare even qualitative predictions 
of KCMs with experimental or numerical results for dynamical 
fluctuations, one has to understand clearly 
in what ensemble and for what dynamics KCMs predictions 
hold.
This certainly highlights the importance of a microscopic 
derivation and more detailed justification of KCMs. 

\section{Numerical results for two molecular glass-formers}
\label{MD}

\subsection{Models and technical details}

In this section we report our numerical
calculations of the dynamic susceptibility $\chi_T(t)$ in two
molecular glass-formers which have been extensively studied in
numerical simulations: a binary Lennard-Jones (LJ) mixture~\cite{KA},
considered as a simple model system for fragile supercooled 
liquids~\cite{hans},
and the Beest, Kramer, and van Santen (BKS)
model, which is a simple description of the strong
glass-former silica~\cite{bks_sim,beest90}. 
For both models we have investigated the 
behavior of the dynamical fluctuations performing 
microcanonical simulations at constant energy, $E$, number of particles, $N$, 
and volume, $V$, by solving Newton's equations of motion~\cite{at}. 
For the LJ
system we have also  simulated two types of 
stochastic dynamics, namely Brownian
and Monte-Carlo dynamics~\cite{at}. 

We follow the dynamical behavior 
of the molecular liquids through the self-intermediate
scattering function, 
\be F_s({\bf k},t) =
\left\langle \frac{1}{N_\alpha} \sum_{j=1}^{N_\alpha} e^{i {\bf k}
\cdot [{\bf r}_j(t) - {\bf r}_j(0)]} \right\rangle, 
\label{self} 
\ee
where the sum in Eq.~(\ref{self}) runs over one of the species of the
considered liquid ($A$ or $B$ in the LJ, Si or O for silica). We
denote by ${f}_s({\bf k}, t)$ the real part of the instantaneous value
of this quantity, so that we have $F_s({\bf k}, t) = \langle
{f}_s({\bf k},t)\rangle$.

As usual, 
the four-point susceptibility, $\chi_4(t)$, quantifies the strength of
the spontaneous fluctuations around the average dynamics by their
variance, \be
\chi_4(t) = N_\alpha \left[  \langle {f}_s^2({\bf k}, t)
\rangle - F_s^2({\bf k}, t) \right]. \label{chi4lj} \ee In
principle, $\chi_4(t)$ in Eq.~(\ref{chi4lj}) retains a dependence on
the scattering vector ${\bf k}$. 
Since the system is isotropic, 
we circularly average (\ref{self}) and
(\ref{chi4lj}) over wavevectors of fixed modulus, 
and we mainly consider results for $|{\bf k}| = 7.21$
for the LJ system, and $|{\bf k}| = 1.7$~\AA$^{-1}$ for the
BKS. These values respectively represent the typical distance
between $A$ particles, and the size of the SiO$_4$ tetrahedra.
Finally, we use finite difference to evaluate 
the temperature derivatives involved in 
\be \chi_T(t) =
\frac{\partial}{\partial T} F_s({\bf k},t).
\ee

We have given an extensive account of the models, numerical details
and parameters used in I. Therefore, we refer 
readers interested in the technical details concerning the 
simulations to I~\cite{BBBKMRI}. Note also that we will neglect,
as it is justified in I, the role density fluctuations on dynamical
correlations. Therefore we will focus
on $\chi_4^{NVT}$ (instead of $\chi_4^{NPT}$) and $\chi_T$ obtained deriving 
with respect to temperature at fixed volume (and not fixed pressure).   

\subsection{Time dependence of dynamic susceptibilities}

\subsubsection{Time behavior of $\chi_T(t)$}

\begin{figure}
\psfig{file=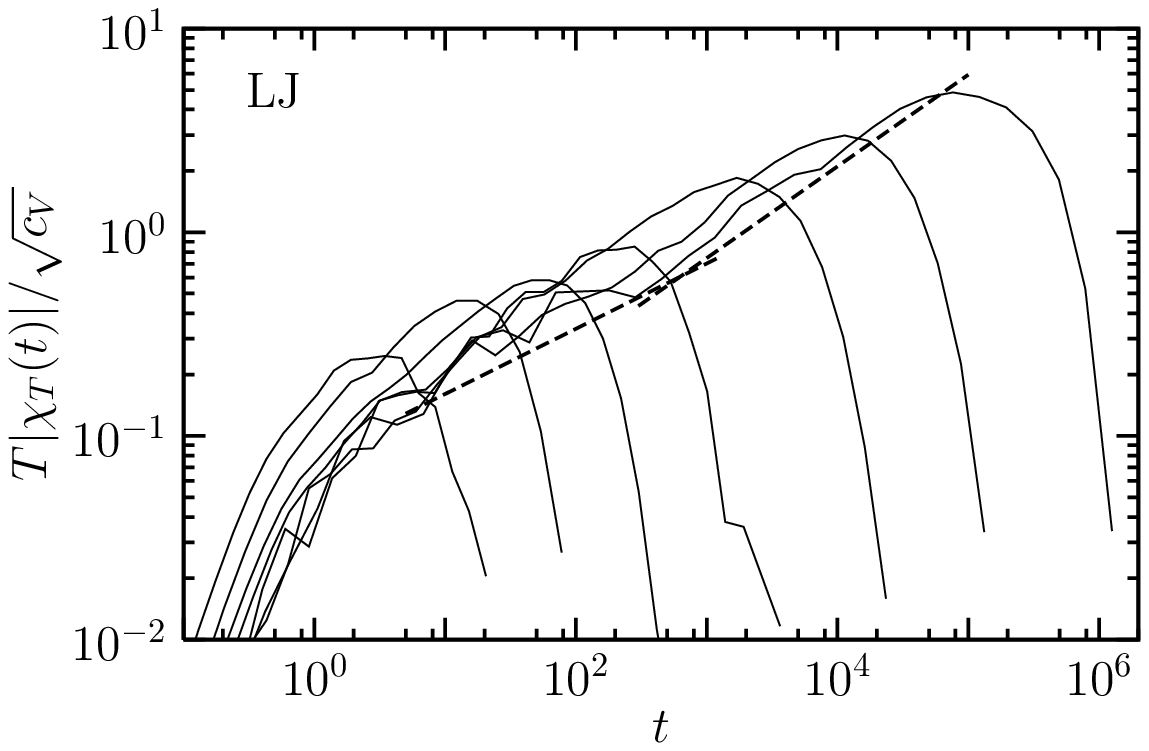,width=8.5cm}
\psfig{file=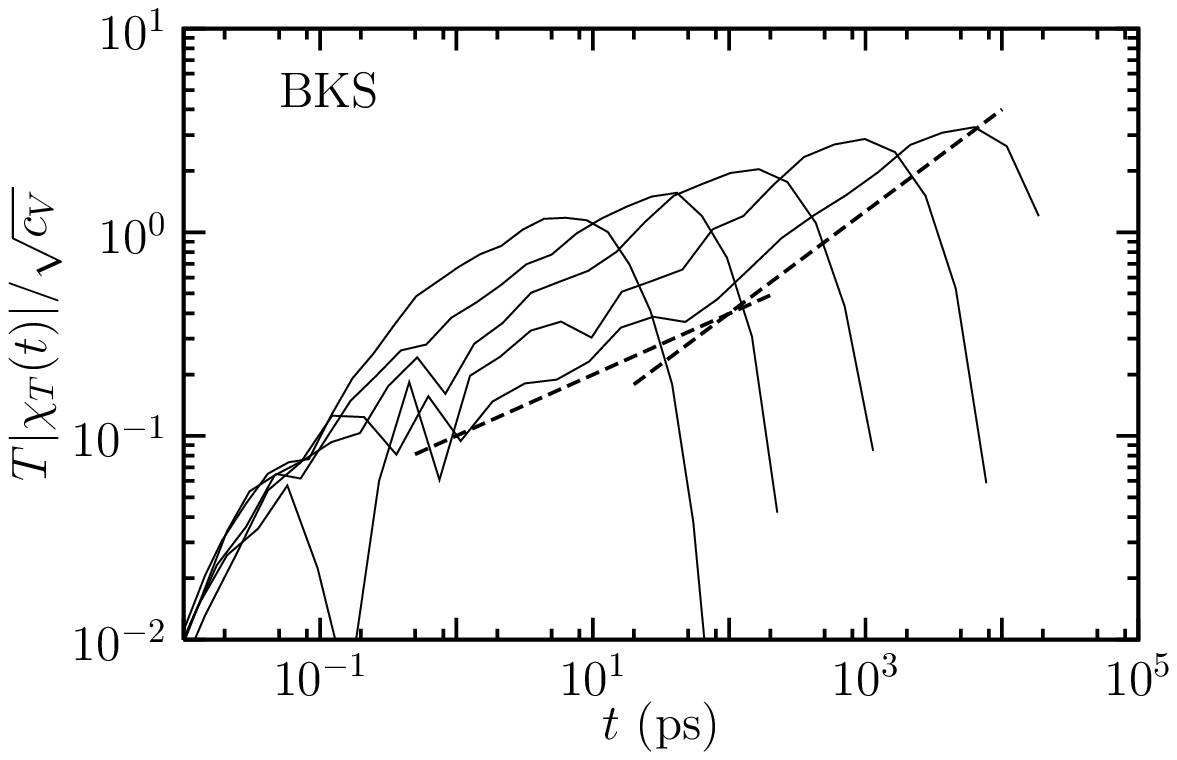,width=8.5cm}
\caption{\label{figBKS1} 
Normalized $\chi_T$ as a function of time for various temperatures
in a binary Lennard-Jones mixture (top) and the BKS model for silica
(bottom), obtained from molecular dynamics numerical simulations. 
LJ: $T=2.0$, 1.0, 0.75, 0.6, 0.5, 0.465, 0.432 from left to right.
BKS $T=6000$, 4650, 4000, 3550,
3200, 3000, and 2730 K from left to right. 
We have taken the absolute value since $\chi_T$ is a negative quantity.
Power law fits of the time dependence are discussed 
in detail in Sec.~\ref{MD}. The value of the exponents at short and long times
are 0.32 and 0.45 in (c), 0.3 and 0.5 in (d).} 
\end{figure}

Our results for the dynamic susceptibilities, $\chi_T(t)$, are presented
in Fig.~\ref{figBKS1} for both the LJ and BKS models. 
For a given temperature, the qualitative 
time dependence of $\chi_T(t)$ observed in 
Fig.~\ref{figBKS1} resembles the one 
already reported for $\chi_4(t)$: $\chi_T(t)$ presents 
a peak for a timescale close to $\tau_\alpha$. 
This is very natural since by definition 
$\chi_T(t=0) = \chi_T (t \to \infty) = 0$, and 
it is for times $t \approx \tau_\alpha$ that the dynamics 
is most sensitive to temperature changes. We have shown 
the quantity $|\chi_T(t)|$ in these figures, as $\chi_T(t)$ is obviously 
a negative quantity: 
raising the temperature makes the dynamics faster, and hence two-time 
correlators smaller, so that $\partial F_s / \partial T < 0$. 

More quantitatively, we expect the 
two-timescale relaxation of the averaged dynamics 
to lead to a complex time behavior of $\chi_T(t)$, 
similar to that predicted for $\chi_4(t)$~\cite{TWBBB}. 
Within MCT, we expect 
(see Sec.~\ref{sectionMCT})
two distinct 
power laws, $\chi_T \sim t^a$ followed
by $\chi_T\sim t^b$, to describe the approach to the maximum
of $\chi_T$, the exponents $a$ and $b$ being already
constrained to the values they take when fitting
the averaged dynamics using MCT. 
From the study of KCMs only the approach to the peak can
be predicted since the short-time dynamics contains no 
clear relaxation towards a plateau due to the coarse-grained nature
of the models~\cite{TWBBB,Ck,nef}. Again, 
a power law approach to the peak is expected.

In Fig.~\ref{figBKS1} we compare our 
numerical results for $\chi_T(t)$ to power law behaviors shown
as dashed lines. On the restricted time window of the simulations
there is obviously some freedom in the fitting procedure so the exponents
we report should be considered as an empirical quantitative description of 
the true time dependence of these functions.
As discussed already in the case of $\chi_4(t)$~\cite{TWBBB}, corrections
to the asymptotic scaling laws derived by theoretical approaches 
should be expected in the reduced time regime of the molecular 
simulations.
In the LJ system we find that the time behavior of $\chi_T(t)$ 
can be described by the exponents 
$a \approx 0.32$ and $b \approx 0.45$ with the tendency that these 
exponents very slowly decrease when $T$ decreases. For the BKS system
we find a similar quality of the fits with $a\approx 0.3$ and 
$b \approx 0.5$ with no systematic dependence in temperature.  

The values of these exponents compare reasonably well with the MCT 
predictions obtained above. For the LJ system, the von-Schweidler
exponent is estimated to be $b \approx 0.51$ from fitting the averaged 
dynamics in the $\beta$-relaxation regime~\cite{KA}, while direct computations
predict $b=0.62$~\cite{nauroth}. 
Both values are close to our finding, $b\approx 0.45$, although 
they both slightly overestimate it.
The exponent $a$ describing the dynamics in the early $\beta$-regime 
was not directly fitted, but using the known relations 
between MCT exponents its value is predicted to be $a=0.29$ 
(for $b=0.51$) and $a=0.32$ (for $b=0.62$). This is again consistent
with our finding for $\chi_T(t)$, $a \approx 0.32$, in this time regime.
From the point of view of MCT, we suggest that focusing on $\chi_T$ 
is a more powerful way to directly measure the exponent $a$ 
(this might be interesting from
an experimental point of view as well).
Finally for BKS, fitting of the average dynamics provides the value $b=0.62$, 
from which $a=0.32$ is deduced from known MCT relations~\cite{bks_sim}. 
These two values again compare relatively well with the time 
behavior found for $\chi_T(t)$, namely $a\approx 0.3$ and $b\approx 0.5$.

\begin{table*}
\begin{tabular}{||l|c|c|c|c|c|c||} \hline\hline
Observable\ \hspace{0.1cm} \   & LJ & BKS  & MCT (LJ) & MCT (BKS) & KCM (1FA) & KCM (East)  \\ \hline
$\theta$ ($\chi_T$)            & 0.33  & 0.14   & 0.43 &  0.43  & 0.25 & $\propto T$ \\ \hline
$\theta$ ($\chi_4^{NVE}$)      & 0.39 &  0.18  & 0.43 &  0.43 & 0.5$^*$ & $\propto T$ \\ \hline
$a$ ($\chi_T$)                 & 0.32 &  0.3  & 0.29-0.32 &  0.32 & NA & NA \\ \hline
$a$ ($\chi_4^{NVE}$)           & 0.37$^1$ & NA   & 0.29-0.32 &  0.32 & NA & NA \\ \hline
$b$ ($\chi_T$)                 & 0.45 &  0.5  & 0.51-0.62 &  0.62  &  1 & $\beta(T) \propto T$ \\ \hline
$b$ ($\chi_4^{NVE}$)           & 0.7 &  0.65-0.85  & 0.51-0.62 &  0.62 & 2$^*$ & $\beta(T) \propto T$ \\ \hline \hline
\end{tabular}
\caption{Summary
of the different results for exponents $\theta, a$ and $b$, describing the peak amplitude and the
time dependence of $T |\chi_T|/\sqrt{c_V}$ and $\chi_4^{NVE}$ (see text). NA: not applicable; $^1$: Obtained from MC dynamics; 
$^*$: Ambiguous --
do KCMs describe $\chi_4^{NVT}$ Newtonian or $\chi_4^{NVT}$ Brownian ($=\chi_4^{NVE}$)?}
\end{table*}

Applying results from KCMs to real liquids, one would predict
the time dependence of $\chi_T(t)$ when approaching the peak to be 
$\chi_T(t) \sim t$ for an Arrhenius liquid modelled by the 
1-spin facilitated model in three dimensions, 
while $\chi_T(t) \sim t^{b(T)}$ is predicted 
for fragile liquids modelled by the East model.
Our numerical results for BKS silica are not consistent with 
the FA model predictions and are, quite 
unexpectedly, more compatible with 
the smaller exponents observed in the fragile East model
reported in Section~\ref{1dkcm}.
The small $b(T)$ exponents 
of the East model compare however well with the behavior of 
$\chi_T(t)$ found in the LJ system. 
In particular, the fact that $b(T)$
decreases with decreasing $T$ is correctly predicted by fragile
KCMs, as opposed to the constant $b$ predicted by MCT. For a summary of 
these results, see Table I.

\subsubsection{Comparison between $\chi_4(t)$ and $\chi_T(t)$}

\begin{figure}
\psfig{file=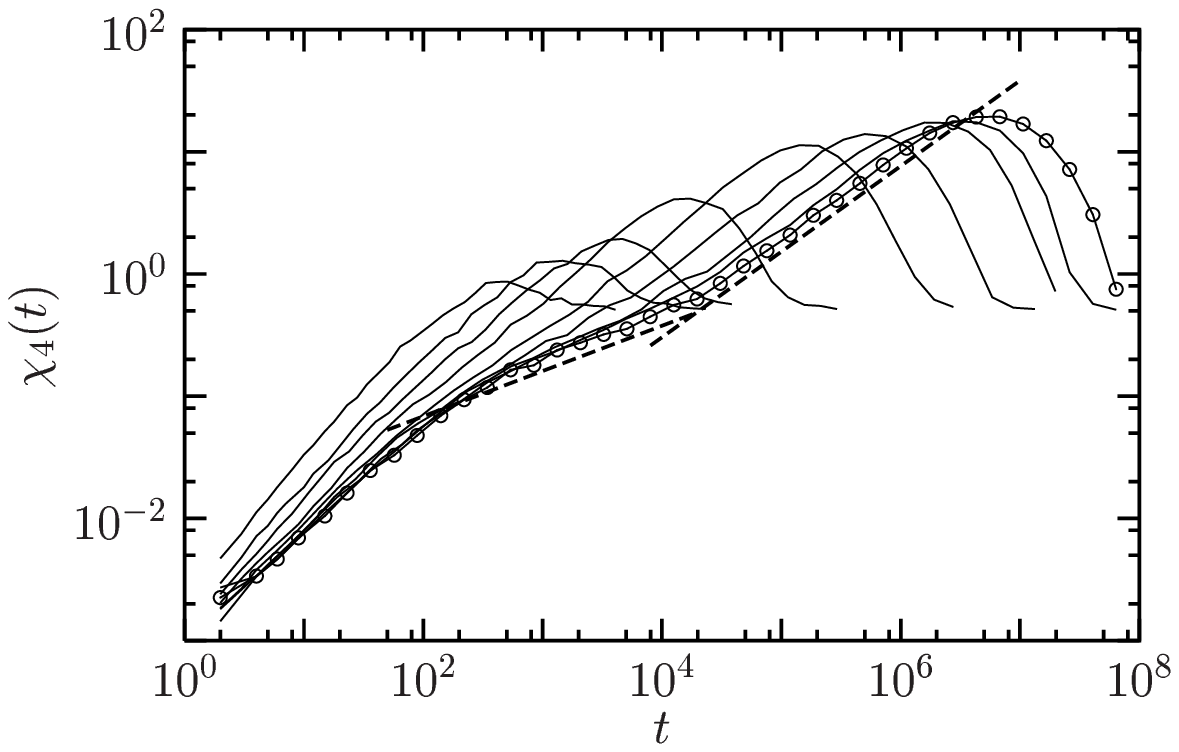,width=8.5cm}
\psfig{file=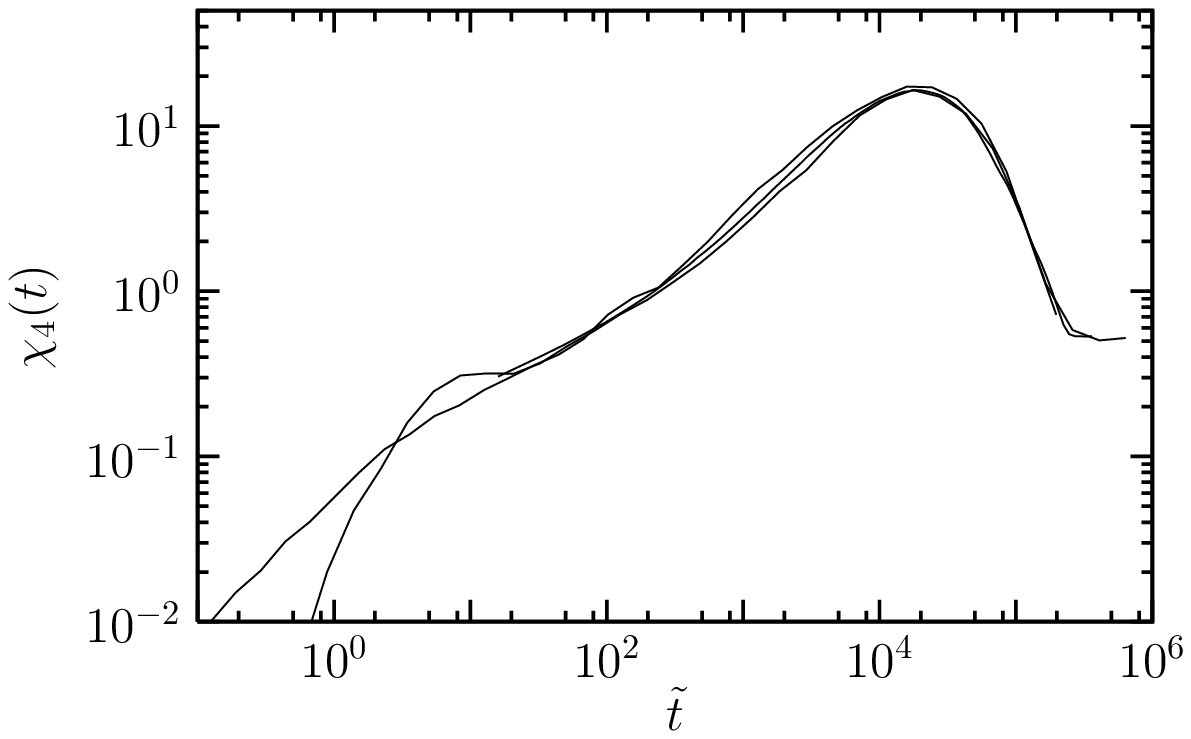,width=8.5cm}
\caption{\label{LJMC1} Top: Four-point susceptibility $\chi_4(t)$
in the binary LJ mixture with Monte-Carlo dynamics
for $T=2.0$, 1.0, 0.75, 0.6, 0.5, 0.47, 0.45 and  0.43 (from left to right),
the lowest temperature being highlighted with open circles. 
Power laws $\chi_4 \sim t^{0.37}$ and $\chi_4 \sim t^{0.7}$
are indicated with dashed lines in the early and late $\beta$-regimes, 
respectively.
Bottom: $\chi_4(t)$ is shown for $T=0.45$ for $NVE$ Newtonian,
Brownian and Monte Carlo dynamics as a function of a rescaled time
chosen so that all $\chi_4$'s overlap near the alpha relation.
We chose $\tilde{t}=t$ for $NVE$ Newtonian dynamics, $\tilde{t}=t/24$ 
for Brownian dynamics, $\tilde{t}=t/100$ for Monte Carlo dynamics.
No rescaling of the vertical axis is performed:
The agreement between the 3 types of dynamics is 
remarkable.}
\end{figure}

It is interesting to compare the exponents found numerically for 
$\chi_T(t)$ to the ones of $\chi_4(t)$ measured 
in the $NVE$ ensemble for 
Newtonian dynamics since theory predicts some relations between 
them.
The latter exponents
were already studied in Ref.~\cite{TWBBB} for the LJ. Numerically
no power law behavior $\chi_4(t)\sim t^a$ is found
in the short-time behavior of $\chi_4(t)$ in the Newtonian
dynamics of both the LJ and BKS systems. This is due to the fact
that thermal vibrations strongly affect the short-time dynamics
of these liquids. Two power-law regimes are however 
clearly observed in the stochastic simulations 
where phonons are either overdamped (Brownian dynamics), 
or absent (Monte-Carlo dynamics). Our Monte-Carlo
results for $\chi_4(t)$ in the LJ are presented in Fig.~\ref{LJMC1} (top)
where we have fitted the early and late $\beta$ regimes with 
two power laws with exponents $a \approx 0.37$ and $b \approx 0.7$, 
respectively. For the BKS we performed Newtonian dynamics simulations only. 
Hence, we only have results on the exponent $b$ from 
$\chi_4$ measurements, which is found 
to increase from 0.65 to 0.85 upon lowering the temperature: this
is an opposite  behavior compared to the LJ where $b$ decreases.
This might suggest a different temperature behavior of $b$ in 
strong and fragile liquids. This trend is 
partly captured by KCMs.

MCT predicts that $\chi_T(t)$ and $\chi_4^{NVE}(t)$ 
have the same critical scaling. 
KCMs predictions are ambiguous so we
follow the numerical results obtained in Sec.~\ref{sectionkcm}, i.e. 
$\chi_4(t) \sim \chi_T^2(t)$.
In both LJ and BKS
systems, the exponent $a$ is the same for both susceptibilities, 
as predicted by MCT. 
The results for $b$ are more difficult to interpret: although 
$b$ for $\chi_4$ is systematically larger than
for $\chi_T$, the ratio between the two exponents is not 2 either, 
so that neither MCT nor KCMs approaches really describe 
this aspect of our numerical results. For a summary
of these results, see Table I. 

What comes nicely out of the simulations, however, is the
fact, predicted on general grounds in I and within MCT above,
that $NVE$ Newtonian, Brownian and Monte Carlo dynamics
display similar time dependences for the dynamic
susceptibility $\chi_4(t)$. This is strikingly illustrated in 
Fig.~\ref{LJMC1} (bottom) which shows $\chi_4(t)$ at a single
temperature, $T=0.45$. The results for the three dynamics almost perfectly 
overlap for timescales larger than the plateau regime 
in $F_s({\bf k},t)$. 

\subsection{Peak amplitude of dynamic fluctuations}

\begin{figure}
\psfig{file=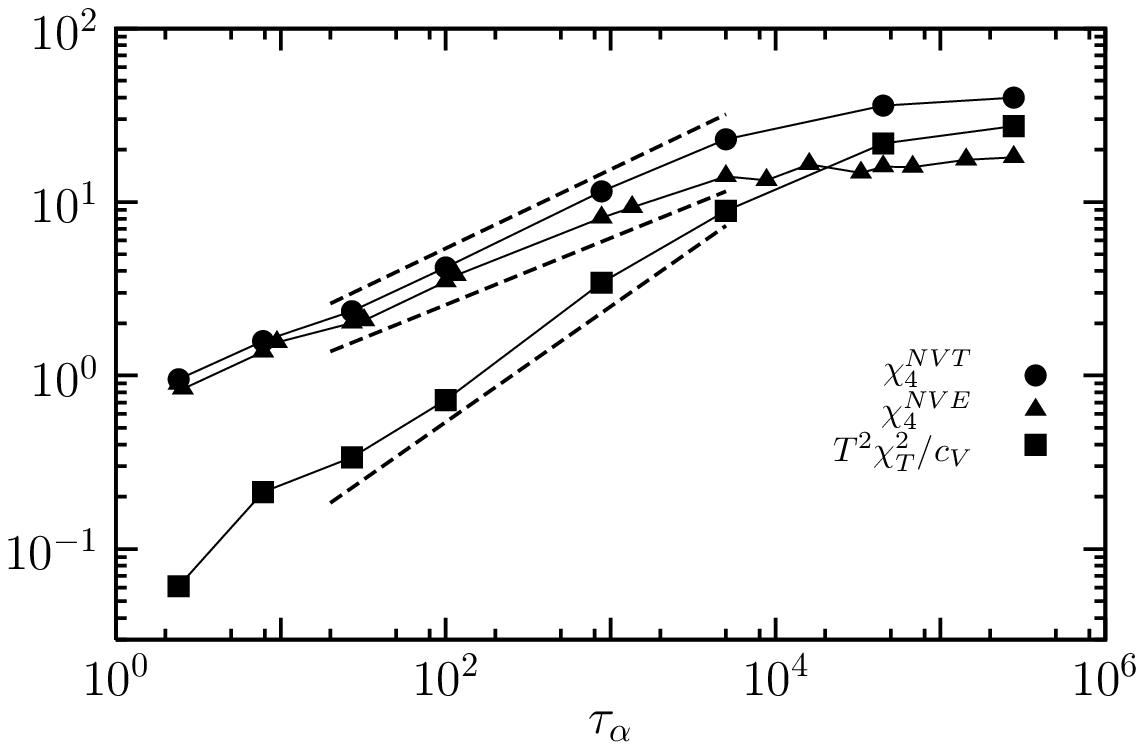,width=8.5cm}
\psfig{file=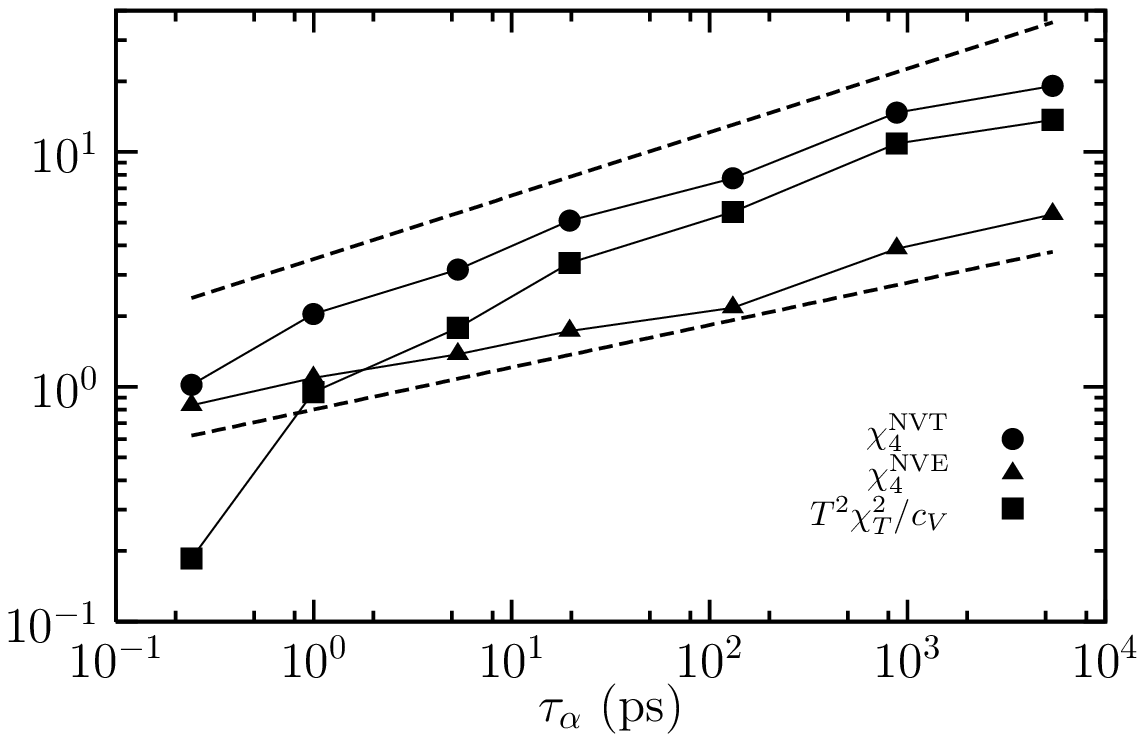,width=8.5cm}
\caption{\label{figLJ3} 
Various susceptibilities in the binary LJ mixture obtained 
from the A particles dynamics (top) and 
the BKS model for silica from the Si ions dynamics (bottom).
Dashed lines indicate power law behavior with exponents
0.46, 0.39 and 0.67 (from top to bottom in the LJ system), 
and 0.27 and 0.18 (from top to bottom in the BKS model).
In all cases, $T^2 \chi_T^2 /c_V$ is smaller than $\chi_4^{NVE}$ 
at high temperature, but increases faster and becomes 
eventually the dominant contribution to $\chi_4^{NVT}$ 
in the relevant low temperature glassy regime.}
\end{figure}

We now focus on the amplitude of the peak observed in the various 
susceptibilities. In Fig.~\ref{figLJ3}, we present our numerical 
results for $\chi_4^{NVE}$, $T^2 \chi_T^2 /c_V$ and 
their sum $\chi_4^{NVT}$ obtained from  
the Newtonian dynamics of both the LJ and BKS models.
When temperature decreases, all peaks shift to larger times 
and track the $\alpha$-relaxation. Simultaneously, their height
increases, revealing increasingly larger dynamic fluctuations as
the glass transition is approached. 

The main observation from the data displayed in 
Fig.~\ref{figLJ3}, already made in Ref.~\cite{science} and in I, 
is that in both LJ and BKS systems
the term  $T^2 \chi_T^2 / c_V$ while being small, $\sim~O(10^{-1})$, 
above the onset temperature of slow dynamics, grows much
faster than $\chi_4^{NVE}$ when the glassy regime 
is entered. As a consequence, there exists a temperature 
below which the temperature derivative contribution to the four-point 
susceptibility $\chi_4^{NVT}$ dominates over that of $\chi_4^{NVE}$.
This crossover is located at 
$T\approx 0.45$ in the LJ system, 
$T \approx 4500$~K  for BKS silica.
Remarkably, the conclusion that $T^2 \chi_T^2 /c_V$ becomes
larger than $\chi_4^{NVE}$ at low temperatures 
holds for both strong and fragile 
glass-formers.
Experimental and theoretical consequences of this observation were 
discussed in Refs.~\cite{science,BBBKMRI}.

Following Ref.~\cite{steve} we have chosen to present the evolution of 
the amplitude of the dynamic susceptibilities  
as a function of $\tau_\alpha$ rather than $T$ because 
it is in this representation that dynamic scaling might emerge. 
For the LJ system we find that all susceptibilities 
can be described by power laws, $\chi \sim \tau_\alpha^{\theta}$, 
in some intermediate, and 
therefore subjectively defined, temperature regime with following 
exponents: $\theta \approx 0.39$ for $\chi_4^{NVE}$, 
$\theta \approx 0.46$ for $\chi_4^{NVT}$ 
and $\theta \approx 0.67$ for $T^2 \chi_T^2 /c_V$.
For the BKS model, we find $\theta \approx 0.27$
for both $\chi_4^{NVT}$ and, in a more restricted time window, 
$T^2 \chi_T^2/c_V$ while we find $\theta \approx 0.18$ for $\chi_4^{NVE}$. 

The theoretical considerations given above show that these exponents 
should be related, within MCT, to the exponent $\gamma$ describing
the divergence of $\tau_\alpha$ close to $T_c$. The prediction 
is that $\theta = 1/\gamma$ for $\chi_4^{NVE}$ and 
$T \chi_T / \sqrt{c_V}$~\footnote{Note that,
within MCT, $T\chi_T/\sqrt{c_V}$ and $\chi_T$
have the same scaling when approaching the critical point.
In real materials, the MCT transition (if any) is avoided and 
temperature dependent prefactors might slightly change 
the observed scaling behavior. 
Here we have chosen to focus on 
$T\chi_T/\sqrt{c_V}$ because it corresponds to a 
properly normalized correlation 
function.} while 
$\theta = 2/\gamma$ for $\chi_4^{NVT}$ and $T^2 \chi_T^2/c_V$.  
Fitting of the relaxation times has shown that $\gamma \approx 2.35$ 
for both LJ and BKS systems, so the exponents $1/\gamma =  0.426$ 
and $2/\gamma = 0.851$ should be observed in Fig.~\ref{figLJ3}. 
The exponent for $\chi_4^{NVE}$ 
is reasonably well 
described by MCT predictions in the LJ system, 
an agreement already reported in Refs.~\cite{BB,szamel} (see Table I). 
The agreement deteriorates somewhat
for $T\chi_T/\sqrt{c_V}$.  The MCT predictions fail however strongly 
in the BKS system, for which the value 0.18 is found instead 
of the expected 0.426 for $\chi_4^{NVE}$ and $\chi_T$, although
in a temperature regime where Arrhenius behavior is already 
observed. No clear power law can be seen in the mode-coupling
regime seen in \cite{bks_sim}.
In principle, the behavior of $\chi_T(t)$ is 
completely tied to the one of the average two-time correlators
already studied in \cite{bks_sim}, but $\chi_T(t)$ provides
a more detailed analysis of the dynamics
with no fitting procedure required. 
Therefore the failure of MCT to capture the behavior 
of $\chi_T(t)$ suggests that MCT, despite the claims
of \cite{bks_sim}, does not satisfactorily describe the 
dynamical behavior of this strong glass-former.

Finally we find that $T^2 \chi_T^2/c_V$ and $\chi_4^{NVT}$
behave somewhat differently in the temperature regime where
power law fits are performed. This is not surprising. 
We have extensively discussed in I the fact that 
simulations are typically performed in the relatively high
temperature regime where both terms contributing to $\chi_4^{NVT}$
are comparable. Since they are predicted to have different scaling 
behaviors, the intermediate value for the exponent $\theta$ reported for 
$\chi_4^{NVT}$ simply results from this crossover.   

The power law regimes we have discussed do not describe the whole 
temperature range studied for the LJ system. For $T \lesssim 0.47$ the growth
of all dynamic susceptibilities with $\tau_\alpha$ 
becomes much slower, perhaps logarithmically slow, but we do not
have a sufficient range of timescales in this low temperature regime 
to draw more quantitative conclusions.
We have moreover checked that this saturation is not the finite
size effect expected if fluctuations are computed in too small 
a system size~\cite{fss}, see~I. Interestingly, no such saturation
can be observed in the BKS system. Therefore we do not know 
how to extrapolate the present numerical results 
towards the glass transition temperature, and compare
our simulations to the result, reported in Ref.~\cite{science}, 
that dynamic susceptibilities have typically the same value 
at $T_g$ for liquids with very different fragilities.  
We can simply state from our results 
that this fragility-independence cannot hold 
at all temperatures since Fig.~\ref{figLJ3} clearly shows that
dynamic susceptibilities grow at different rates in different systems.
We are currently investigating this point in more 
detail~\cite{exppaperinprep}.

The saturation of the LJ dynamic susceptibilities observed at low $T$ 
seems consistent with 
the theoretical 
expectation~\cite{Wolynes,TWBBB,nef,arrow,rfot,Gilles,droplet}, and the 
experimental confirmation~\cite{science,exp0,mark,encoremark} 
that dynamic fluctuations and lengthscales grow very slowly 
when $T$ is decreased towards $T_g$. From the fragile 
KCMs perspective, one would for instance expect 
that $\chi_4 \sim \tau_\alpha^{\theta(T)}$ with an exponent 
$\theta(T)$ which decreases
linearly with $T$~\cite{nef,arrow}, 
while logarithmic growth, $\chi_4 \sim (\log \tau_\alpha)^\psi$, 
is predicted by activation based theories ~\cite{rfot,Wolynes,droplet}.

\section{Conclusions and perspectives}
\label{conclusion}

This paper describes the second part of our investigations 
of dynamical susceptibilities started in I~\cite{BBBKMRI}.
In this second work we have illustrated the 
general conclusions of I  by making explicit the predictions 
of MCT and KCMs concerning spontaneous
dynamical fluctuations (encoded in $\chi_4(t)$) and induced one (given 
by $\chi_x(t)$). These theories predict the detailed dependence of these two
quantities both as a function of time and of temperature (or density). As 
discussed in I, special care must be devoted to the choice of statistical 
ensemble and microscopic dynamics, with the rather spectacular prediction 
of MCT that $\chi_4(t)$ should coincide (or at least display the same scaling) 
for Newtonian dynamics in the $NVE$ 
ensemble and for Brownian dynamics in the $NVT$ ensemble, but differ from the
result for Newtonian dynamics in the $NVT$ ensemble. 
The predictions coming from
KCMs are much less clear about this particular point, 
since there is some intrinsic ambiguity about which 
ensemble and which dynamics these models are supposed to describe. 

We have compared these predictions with numerical simulations of 
models of supercooled
liquids. 
Overall, as shown in Table I, MCT fares reasonably 
well at accounting for the 
detailed shape of $\chi_4(t)$
and $\chi_T(t)$ of the Lennard-Jones system, in a restricted 
temperature region where MCT can be applied.
As for the values of the exponents, our aim was to present a
rather qualitative comparison focusing more on compatibility than 
on precise tests, which are beyond the scope of this work,
and probably of 
MCT as well.
Quite remarkably, the exponents used to 
fit these higher order 
correlations are indeed compatible with those measured on two-point correlation
functions, with quantitative variations that can perhaps be attributed
to preasymptotic effects. 
Furthermore, the predicted ensemble dependence of 
these quantities is 
very clearly highlighted by our numerical results. We have also 
shown that the wave-vector 
dependence of $\chi_4(t)$ can be qualitatively
accounted for within MCT. On the 
other hand, the features 
of the dynamical susceptibility of the BKS model for the strong 
silica glass are not quantitatively well explained by MCT.
Similarly KCMs fail to describe quantitatively
the results obtained in the BKS model, but  
the systematic temperature dependence of the exponents 
describing $\chi_4(t)$ appears somewhat 
natural from this perspective.

Among open problems, we should primarily emphasize the major 
problem of extending MCT to 
allow for activated events.
A detailed prediction of $\chi_4(t)$ and of the geometry and 
exponents of dynamically correlated regions
in the deeply supercooled region would be important to compare 
with future experiments (see~\cite{TWBBB,rfotfrac,BBMR}
for preliminary elements in that direction). The generalization 
of these predictions to the aging regime would also 
certainly be relevant to analyze the cooperative dynamics of 
deeply quenched glasses.

\acknowledgments
We thank L. Cipelletti, S. Franz,
F. Ladieu, A. Lef\`evre, D. L'H{\^o}te, G. Szamel, and G. Tarjus for discussions. 
D.R.R. and K.M. acknowledge support from the NSF (NSF CHE-0134969). 
G.B. is partially supported by EU contract HPRN-CT-2002-00307 (DYGLAGEMEM).
K.M. would like to thank J. D. Eaves for his help on 
the development of efficient numerical codes.

\renewcommand{\thesection}{Appendix \Alph{section}}
\renewcommand{\thesubsection}{\Alph{section}.\arabic{subsection}}
\renewcommand{\theequation}{\Alph{section}.\arabic{equation}}  
\setcounter{section}{0}

\section{Dynamic susceptibilities in the $p$-spin model}
\label{appendixkuni}
\setcounter{equation}{0}

\subsection{General discussion and results}
\label{pspin}

Much intuition concerning dynamic heterogeneity has been gleaned from the 
study of mean-field spin-glasses.  In particular, Franz and Parisi first 
pointed out that a quantity analogous to 
$\chi_{4}(t)$, which can be computed exactly in 
mean-field $p$-spin models, should show non-trivial features~\cite{FP}, 
which prompted the study of dynamic fluctuations in simulations of atomistic 
glass-forming liquids~\cite{silvio2}. The growth of 
a dynamic susceptibility in this 
model was properly interpreted in terms of a growing dynamical length 
scale, which diverges at $T_{c}$.  The 
same scenario, complete with a temporal behavior of $\chi_{4}(t)$ 
identical to that in the $p$-spin models, exists in mean-field models that 
have no underlying thermodynamic critical point~\cite{ioffe1,sellitto}.
It should also be noted that this scenario is perhaps more general than 
appreciated, since it appears to also exist in models on compact 
lattices with no quenched disorder and short-ranged interactions, at least  
in the limit of large dimensionality~\cite{ioffe2}, 
and models with long-ranged, Kac-like 
interactions~\cite{KacSchmalian,KacFranz}.

Applying the above diagrammatic analysis  to $p$-spin models for which 
no conserved quantities exist, one finds, 
in agreement with BB, that 
$\chi_{4}(t)$ is determined by ladder diagrams only. Hence, its
critical behavior has to be the same as the dynamical response $\chi_T(t)$
and is given by Eq.~(\ref{dr1}).
Similarly the susceptiblity $\chi_{\rm FP}(t)$ introduced
by Franz and Parisi is found to follow the same scaling
behaviour. As discussed below, Franz 
and Parisi~\cite{FP} 
study the quantity $\chi_{\rm FP}(t) =dC(t) / 
d \epsilon$, where $C(t) = \langle s_i(t) s_i(0)
\rangle$ and $\epsilon$ is an infinitesimal field coupling 
the system's configuration at time $t$ to its initial 
state at time $0$. Using linear response theory
they argue that $dC(t)/d\epsilon$ and $\chi_4(t)$ are equal.
We find instead that $dC(t)/d\epsilon$ is equal to the sum of $\chi_4(t)$
and another non-vanishing contribution.
However $dC(t)/d\epsilon=\chi_{FP}(t)=N^{-1}
\sum_{ij} \int_0^t dt'  
\langle s_i(t) s_i(0) s_j(t') \hat{s_j}(t'^+) \rangle$,
where $\hat s_i(t)$ are the response field.
Hence, it is given by ladder diagrams 
similar to the ones contributing to
$\chi_4(t)$.
Thus we expect that 
$\chi_{FP}(t)$ and $\chi_4(t)$ behave similarly close to
the critical point.

In the following, we shall present a careful numerical comparison between
the dynamic susceptibility $\chi_{\rm FP}(t)$ 
and $\chi_T(t)$ integrating the
integro-differential equations derived in ~\cite{FP} for 
$p$-spins models. This comparison decisively confirms 
the previous analytical results. A much smaller time 
window was studied in \cite{BB}, and it was not 
clear that asymptotic regimes had been observed.

One technical difficulty is that it is numerically 
difficult to calculate $\chi_{\rm FP}(t)$ very close to $T_{c}$.  
Here, we modify the method developed by Kim and Latz for the
aging $p$-spin model~\cite{KL} to accurately integrate the equations on  
$\chi_{\rm FP}(t)$ 
derived in Ref.~\cite{FP} much closer to $T_{c}$ than has been
reported in previous work.  
The dynamical equations are presented in \ref{kunidyneq}, while the
details of the methodology are outlined in \ref{kunimethod}.
In the $p$-spin case, one can use an alternative way to compute 
$\chi_{\rm FP}(t)$ 
based on power counting in $N^{-1}$, the inverse number of spins. 
This provides
a complementary way to show that dynamical fluctuations are 
indeed given by ladder diagrams.
\begin{figure}
\psfig{file=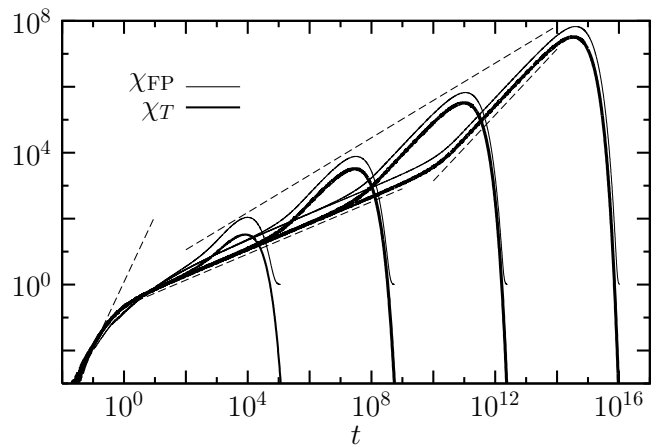,width=8.5cm}
\caption{\label{figMCT} Time dependence of the 
dynamic susceptibilities $\chi_T(t)$ (thick lines) and $\chi_{\rm FP}(t)$ 
(thin lines) in the $p=3$ mean-field $p$-spin model for 
temperatures approaching $T_c$ from above.
Note the wide range of timescales covered in this graph.
From left to right, $(T-T_c)/T_c = 10^{-2}$, $10^{-4}$, 
$10^{-6}$, $10^{-8}$.
The asymptotic power law regimes are shown as dashed lines. The 
susceptibilities grow as $t^2$, $t^a$ and $t^b$ in the 
microscopic, early and late beta 
regimes, while the height of the maxima scale with 
their as $\chi^\star \sim \tau^{1/\gamma}$. 
For $p=3$, one has $a=0.395$, $b=1$ and $\gamma=1.765$.} 
\end{figure}

Let us now present our numerical results. 
In Fig.~\ref{figMCT}, we show a comparison of 
$\chi_{\rm FP}(t)$ and $\chi_{T}(t)$ for various temperatures approaching
$T_{c}$ from above. Clearly, $\chi_{T}(t)$ 
is remarkably similar to $\chi_{\rm FP}(t)$ in this regime, exhibiting a well 
defined regime at short times that grows as a power-law with the 
critical mode-coupling exponent $a=0.395$, and a well-defined power-law at 
later times that grows with the von Schweidler exponent $b=1$.  
Note also that the height of the peak scales 
as $\tau^{1/\gamma}$ (where $\tau$ is the relaxation time) for
both functions, as predicted.
When the transition temperature is 
approached from the non-ergodic phase, only the first regime of 
slow growth with the exponent $a$ can be observed (not shown). 
These results represents a useful benchmark for the comparison 
with real liquids. Indeed, as 
presented in Fig.~\ref{LJMC1}, $\chi_{4}(t)$ for 
Monte-Carlo dynamics in a binary Lennard-Jones mixture 
(where vibrational modes that may obscure the exponent $a$ are
absent) 
shows features strikingly similar to 
those of the $p$-spin model, complete with a reasonably defined regimes 
showing both $a$ and $b$ exponents close to $T_{c}$. 

\subsection{Exact dynamical Equations}
\label{kunidyneq}

Following Franz and Parisi~\cite{FP},
we consider the dynamic of a perturbed  
$p=3$ spherical $p$-spin model evolving with the Hamiltonian 
$H_{\mbox{\scriptsize tot}}(S) = H(S) - \epsilon \hat{C}({S}, {S}_{0})$,
where ${S}_t$ is the spin state at time $t$, 
$\hat{C}({S}, {S}^{\prime})\equiv N^{-1}\sum_{i}S_iS_i^{\prime}$ 
is the overlap function, 
and  $H(S)= \sum_{i < j< k}J_{ijk}S_{i}S_{j}S_{k}$ is the unperturbed
$p$-spin Hamiltonian. The Franz-Parisi 
susceptibility is defined as the linear response of
the two-point correlation function evaluated in the presence of the
perturbation,  
$C_{\epsilon}(t,0)
\equiv 
\langle \hat{C}({S}_t, {S}_0) \rangle_{\epsilon}$,
as 
\begin{equation}
\chi_{\rm FP}(t) = \frac{\partial C_{\epsilon}(t,0)}{\partial \epsilon}. 
\end{equation}
The equations of motion for $C_{\epsilon}(t,t^{\prime})$ 
and the associated response function $G_{\epsilon}(t,t^{\prime})$ 
are derived using a
standard MSR-approach~\cite{FP,Zinn-Justin}. 
\begin{widetext}
\begin{equation}\label{43}
\left\{
\begin{aligned}
\frac{\partial C_{\epsilon}(t,t^{\prime})}{\partial t}
=
&
-\mu(t)C_{\epsilon}(t,t^{\prime})
 +\int_{0}^{t}\!\!\mbox{d} s~ 
f^{\prime\prime}(C_{\epsilon}(t,s))G_{\epsilon}(t,s)C_{\epsilon}(s,t^{\prime})
%\\
%&
 +\int_{0}^{t^{\prime}}\!\!\mbox{d} s~ 
 f^{\prime}(C_{\epsilon}(t,s))G_{\epsilon}(t^{\prime},s)
\\
&
 +\beta f^{\prime}(C_{\epsilon}(t,0))C_{\epsilon}(t^{\prime},0)
 +\epsilon C_{\epsilon}(t^{\prime},0), 
\\
\pdif{G_{\epsilon}(t,t^{\prime})}{t}
=
&
-\mu(t)G_{\epsilon}(t,t^{\prime})
 +\int_{t^{\prime}}^{t}\!\!\mbox{d} s~ 
f^{\prime\prime}(C_{\epsilon}(t,s))G_{\epsilon}(t,s)G_{\epsilon}(s,t^{\prime}) 
\end{aligned}
\right.
\end{equation}
\end{widetext}
with the damping coefficient
\begin{equation}
\begin{aligned}
&
\mu(t) 
=
T +\epsilon C_{\epsilon}(t,0) 
  +\beta f^{\prime}(C_{\epsilon}(t,0))C_{\epsilon}(t,0)
\\
&
+ \int_{0}^{t}\!\!\mbox{d} s~ 
 \left\{
 f^{\prime\prime}(C_{\epsilon}(t,s))G_{\epsilon}(t,s)C_{\epsilon}(t,s)
 +f^{\prime}(C_{\epsilon}(t,s))G_{\epsilon}(t,s)
 \right\}
\end{aligned}
\end{equation}
and $f(x)= x^3/2$.
We have numerically solved these equations using the 
method described below.
In the limit of $\epsilon \rightarrow 0$, we retrieve 
the equation of motion for the stationary state;
\begin{equation}
\begin{aligned}
\frac{\partial C(t)}{\partial t}
=
&
- TC(t) + \frac{1}{2}\int_{0}^{t}\!\!\mbox{d} s~ 
C^2(t-s)\pdif{C(s)}{s}, 
\end{aligned}
\end{equation}
where $C(t) = C_{\epsilon=0}(t, 0)$.

The temperature derivative 
$\chi_{T}(t)= \partial C(t)/\partial T$
is evaluated by simple numerical differentiation of $C(t)$
with finely spaced temperature points. 

\subsection{Numerical algorithm}
\label{kunimethod}

In the following, we elucidate the technical detail to solve Eq.~(\ref{43}). 
This is a natural generalization 
of an efficient algorithm to solve equilibrium mode-coupling equation 
developed by Fuchs {\it et al.}~\cite{fuchs1991} to nonstationary systems. 
The method given here can be also applied for the aging
dynamics~\cite{KL}. 

First, we shall introduce a new quantity, 
$Q_{\epsilon}(t,t^{\prime})$ by 
\begin{equation}
Q_{\epsilon}(t,t^{\prime}) \equiv 1 - C_{\epsilon}(t,t^{\prime}) 
- \int_{t'}^{t}\!\!{\mbox{d}} s~G_{\epsilon}(t, s), 
\end{equation}
where the subscript $\epsilon$ has been omitted for simplification.
This function monitors the degree of violation of 
the fluctuation-dissipation theorem. 
With this new function, the MCT equation, Eq.~(\ref{43}) can be 
rewritten as 
\begin{widetext}
\begin{equation}
\begin{aligned}
&
\left\{ 
\begin{aligned}
\pdif{C_{\epsilon}(t,t^{\prime})}{t}
=
&
- \mu^{\prime}(t)C_{\epsilon}(t,t^{\prime})
-\int_{t^{\prime}}^{t}\!\!{\mbox{d}} s~ 
\left[
 f^{\prime}(t,s)\pdif{C_{\epsilon}(s, t^{\prime})}{s}
-f^{\prime\prime}(t,s)\pdif{Q_{\epsilon}(t,s)}{s}C_{\epsilon}(s,t^{\prime})
\right]
+P_{\epsilon}(t, t'),
\\
\pdif{Q_{\epsilon}(t,t^{\prime})}{t}
=
&
-1+\mu^{\prime}(t)
-\mu^{\prime}(t)Q_{\epsilon}(t,t^{\prime})
%\\
%&
-\int_{t^{\prime}}^{t}\!\!{\mbox{d}} s~ 
\left[
 f^{\prime}(t,s)\pdif{Q_{\epsilon}(s, t^{\prime})}{s}  
+f^{\prime\prime}(t,s)\pdif{Q_{\epsilon}(t,s)}{s}
\left\{ 1 - Q_{\epsilon}(s,t^{\prime})\right\}
\right]
-P_{\epsilon}(t, t')
\end{aligned}
\right.
\end{aligned}
\label{eq:appendix.FP3}
\end{equation}
\end{widetext}
with $\mu^{\prime}(t)= 1 +  P_{\epsilon}(t, t)$ and
\begin{equation}
\begin{aligned}
&
P_{\epsilon}(t, t')
= 
\epsilon C_{\epsilon}(t^{\prime},0), 
\\
&
 +
\int_{0}^{t^{\prime}}\!\!{\mbox{d}} s~ 
\left[
 f^{\prime}(t,s)\pdif{Q_{\epsilon}(t^{\prime},s)}{s}
+f^{\prime\prime}(t,s)\pdif{Q_{\epsilon}(t,s)}{s}C_{\epsilon}(t^{\prime},s)
\right], 
\end{aligned}
\label{eq:appendix.defofP}
\end{equation}
where $f(t, t')\equiv f(C_{\epsilon}(t,t'))$.  
In the above expression, the temperature $T$ was absorbed to time, so
that all quantities in the equations are dimensionless. 
Integration of Eq.~(\ref{eq:appendix.FP3}) can be implemented by
discretizing the two dimensional plane of the times $(t, t')$ with
$t \geq t'$
into a
cubic lattice of the grid size $\delta$ . 
Note that Eqs.~(\ref{eq:appendix.FP3}, \ref{eq:appendix.defofP}) 
consist of four types of time integrals; 
\begin{equation}
\left\{
\begin{aligned}
&
I^{(1)}(t, t')
= \int_{t'}^{t}\!\!{\mbox{d}} s~ A(t, s)\pdif{B(s,t')}{s},
\\
&
I^{(2)}(t, t')
= \int_{t'}^{t}\!\!{\mbox{d}} s~ A(t, s)\pdif{B(t,s)}{s}C(s,t'),
\\
&
I^{(3)}(t, t')
= \int_{0}^{t'}\!\!{\mbox{d}} s~ A(t, s)\pdif{B(t',s)}{s}, 
\\
&
I^{(4)}(t, t')
= \int_{0}^{t'}\!\!{\mbox{d}} s~ A(t, s)\pdif{B(t,s)}{s}C(t',s).
\end{aligned} 
\right.
\end{equation}
These integrals are evaluated by discretizing the time 
as $t_i= i\delta$ and slicing into pieces as follows. 
$I^{(1)}(t=t_i, t'=t_j)\equiv I^{(1)}_{ij}$ $(i > j)$, for example, 
is written as 
\begin{equation}
\begin{aligned}
I^{(1)}_{ij} 
= 
&
\int_{t_m}^{t_i}\!\!{\mbox{d}} s~ A(t_i, s)\pdif{B(s,t_j)}{s}
+
\int_{t_j}^{t_m}\!\!{\mbox{d}} s~ A(t_i, s)\pdif{B(s,t_j)}{s}
\\
= 
&
A_{i,m}B_{m,j}-A_{i,j}B_{j,j}
+ 
\sum_{l=m+1}^{i}
\int_{t_{l-1}}^{t_l}\!\!{\mbox{d}} s~ A(t_i, s)\pdif{B(s,t_j)}{s}
\\
&
-
\sum_{l=j+1}^{m}
\int_{t_{l-1}}^{t_l}\!\!{\mbox{d}} s~ \pdif{A(t_i, s)}{s}B(s,t_j), 
\end{aligned}
\end{equation}
where $m = [(i-j)/2]$ is the integer closest to but smaller than 
$(i-j)/2$.
Using an approximation, 
\begin{equation}
\int_{t_{1}}^{t_2}\!\!{\mbox{d}} s~ \pdif{A(s)}{s}B(s)
\approx
\{ A(t_2)-A(t_1) \}\times \frac{1}{\delta}\int_{t_{1}}^{t_2}\!\!{\mbox{d}} s~ B(s),
\end{equation}
which is exact up to ${\cal O}(\delta^2)$~\cite{fuchs1991}, 
we arrive at 
\begin{equation}
\begin{aligned}
&
I^{(1)}_{ij} 
= 
A_{i,m}B_{m,j}-A_{i,j}B_{j,j}
\\
&
+\sum_{l=m+1}^{i}
(B_{l,j}-B_{l-1,j})dA^{(v)}_{i,l}
-\sum_{l=j+1}^{m}
(A_{i,l}-A_{i,l-1})dB^{(h)}_{l,j}, 
\end{aligned}
\end{equation}
where 
\begin{equation}
\begin{aligned}
&
dA^{(h)}_{ij}= \frac{1}{\delta}\int_{t_{i-1}}^{t_i}\!\!{\mbox{d}} s~ A(s, t_j), 
\\
&
dA^{(v)}_{ij}
= \frac{1}{\delta}\int_{t_{j-1}}^{t_j}\!\!{\mbox{d}} s~ A(t_i, s)
\end{aligned}
\end{equation}
are the integrals over the horizontal and vertical lattice bond,
respectively (we refer to them as bond integrals).  
Likewise, other integrals can be approximated as follows. 
\begin{equation}
\begin{aligned}
I^{(2)}_{ij} 
= 
&
\sum_{l=m+1}^{i}\frac{1}{2}dA^{(v)}_{i,l}(B_{i,l}-B_{i,l-1})(C_{l,j}+C_{l-1,j})
\\
&
+ 
\sum_{l=j+1}^{m}\frac{1}{2}(A_{i,l}+A_{i,l-1})(B_{i,l}-B_{i,l-1})dC^{(h)}_{l,j},
\\
I^{(3)}_{ij} 
= 
&
A_{ij}B_{jj}- A_{i,0}B_{j,0}
- \sum_{l=1}^{j}(A_{i,l}-A_{i,l-1})dB^{(v)}_{j,l},
\\
I^{(4)}_{ij} 
= 
&
\sum_{l=1}^{j}\frac{1}{2}(A_{i,l}+A_{i,l-1})(B_{i,l}-B_{i,l-1})dC^{(v)}_{j,l}. 
\end{aligned}
\end{equation}
With this discretization, the nonlinear integro-differential equation, 
Eq.~(\ref{eq:appendix.FP3}), can be written in a form of 
a simultaneous nonlinear equation as
\begin{equation}
{\bf V}_{i} = {\bf M}_{i} \cdot {\bf F}_{i}({\bf V}_{i}) + {\bf N}_{i}, 
\label{eq:appendix.SCequation}
\end{equation} 
where 
${\bf V}_{i}=$ $(C_{i0},\cdots, C_{ii},  Q_{i0}, \cdots, Q_{ii})$ 
and 
${\bf F}_{i}({\bf V}_{i})$
$=$
$(f^{\prime}(C_{i0}),\cdots, f^{\prime}(C_{ii}),$
$f^{\prime\prime}(C_{i0}),\cdots, f^{\prime\prime}(C_{ii}))$
are $(2i+2)$-dimensional vectors.  
The matrix ${\bf M}_{i}$ and the vector ${\bf N}_{i}$ 
are functions of the friction coefficient $\mu^{\prime}$, 
the vectors at the earlier  times $({\bf V}_{l}, {\bf F}_{l})$ with $l < i$, 
and a set of bond integrals ${\bf W}=$
($dC^{(h)}$, $dC^{(v)}$, $dQ^{(h)}$, $dQ^{(v)}$, 
$df^{\prime(h)}$, $df^{\prime(v)}$, 
$df^{\prime\prime(h)}$, $df^{\prime\prime(v)}$).
Equation (\ref{eq:appendix.SCequation}) can be solved self-consistently 
using the following procedure. 
\begin{enumerate}
 \item 
       First, prepare the array of exact ${\bf V}_{i}$, ${\bf F}_{i}$, and 
       ${\bf W}$ for $0 \leq j \leq i \leq N_t/2$ with 
       a very small time grid $\delta$ such that $N_t \delta \ll 1$ by short
       time expansion of Eq.~(\ref{eq:appendix.FP3}). 

 \item 
       For $i = N_t/2+1$ and for $j$ very close to but smaller than $i$, 
       we import the values of the previous time, expecting the short time
       dynamics at $(i-j)\delta \ll 1$ is not affected by the perturbed
       field or by aging. 
       More specifically, we choose an integer 
       $N_{\mbox{\scriptsize short}} \ll N_t/2$ and assign the values 
       $C_{i,j} = C_{i-1,j-1}$, $dC^{(h)}_{i,j} = dC^{(h)}_{i-1,j-1}$ 
       and so forth for $i-N_{\mbox{\scriptsize short}} \leq j \leq i$. 

 \item
       For $i = N_t/2+1$ and for $0 \leq j < i-N_{\mbox{\scriptsize short}}$,
       we solve Eq.~(\ref{eq:appendix.SCequation}) 
       self-consistently by iteration.
       The iteration is done by choosing the initial array as
       ${\bf V}_{i}={\bf V}_{i-1}$. 
       The bond integrals are calculated using  
       \begin{equation}
        \left\{
        \begin{aligned}
         &
         dA^{(h)}_{i,j} 
         = \frac{\delta}{12}\left( -A_{i-2,j}+8A_{i-1,j}+5A_{i,j}  \right),
         \\
         &
         dA^{(v)}_{i,j} 
          = \frac{\delta}{12}\left( -A_{i,j+2}+8A_{i,j+1}+5A_{i,j}  \right).
        \end{aligned}
        \right.
        \label{eq:appendix.simpson}
       \end{equation}
       At every iterations of Eq.~(\ref{eq:appendix.SCequation}) 
       for ${\bf V}_{i}$, all elements of the bond 
       integrals $dA^{(h, v)}_{i,j}$ 
       and, thus ${\bf M}$ and ${\bf N}$, are updated using
       Eq.~(\ref{eq:appendix.simpson}).   

 \item
       Keep the procedure 2 and 3 for $N_t/2 \leq i \leq N_t$. \\

 \item
       Once all solution for $0 \leq i \leq N_t$ are obtained, 
       we decimate the number of variables by half in order to 
       save the memory space to explore further for the longer time. 
       We discard half variables and renew all 
       variables by the following rules;
       For ${\bf V}=(C, Q)$, 
       \begin{equation}
        \begin{aligned}
         &
         V_{2i, 2j}   \rightarrow V_{i, j}.  
        \end{aligned}
       \end{equation}
       For bond integrals,
       \begin{equation}
        \left\{
        \begin{aligned}
         &
         \frac{1}{2}\left( dA^{(h)}_{2i,2j} + dA^{(h)}_{2i-1,2j}\right)  
         \rightarrow dA^{(h)}_{i,j}
         \\
         & 
         \frac{1}{2}\left( dA^{(v)}_{2i,2j} + dA^{(v)}_{2i,2j-1}\right)  
         \rightarrow dA^{(v)}_{i,j}. 
        \end{aligned}
        \right.
       \end{equation}
       Then, the time grid is doubled. 
       \begin{equation}
        2\delta \rightarrow \delta.
       \end{equation}
 \item
       Repeat the procedures 2-5 with the doubled grid size. 
\end{enumerate}
We have checked that, 
in order to obtain a stable result up to the order of $t=10^{16}$ as
shown in the present work, 
we need the number of grid of $N_t=1024$ 
and $N_{\mbox{\scriptsize short}}=32$, starting 
the initial grid size of $\delta=10^{-10}$.

\end{document}